\documentclass[nonacm, sigconf]{acmart}
\usepackage{times}
\usepackage{multirow}
\usepackage[nameinlink, capitalise, noabbrev]{cleveref}

\renewcommand\footnotetextcopyrightpermission[1]{} 
\usepackage[skip=0pt]{subcaption}
\usepackage{soul}
\usepackage[export]{adjustbox}
\date{}

\def\BibTeX{{\rm B\kern-.05em{\sc i\kern-.025em b}\kern-.08emT\kern-.1667em\lower.7ex\hbox{E}\kern-.125emX}}

\usepackage{amsmath}

\usepackage[boxruled,linesnumbered,noend]{algorithm2e}

\renewcommand{\textrightarrow}{$\rightarrow$}
\begin{document}

\title{HyperSched: Dynamic Resource Reallocation for Model Development on a Deadline}

%
\author{Richard Liaw}
\affiliation{
    \institution{UC Berkeley}
}

\author{Romil Bhardwaj}
\affiliation{
    \institution{UC Berkeley}
}

\author{Lisa Dunlap}
\affiliation{
    \institution{UC Berkeley}
}

\author{Yitian Zou}
\affiliation{
    \institution{UC Berkeley}
}

\author{Joseph E. Gonzalez}
\affiliation{
    \institution{UC Berkeley}
}

\author{Ion Stoica}
\affiliation{
    \institution{UC Berkeley}
}

\author{Alexey Tumanov}
\affiliation{
    \institution{Georgia Tech}
}

\newcommand{\hypersched}[1]{\textit{HyperSched}}
\newcommand{\Hypersched}[1]{\textit{HyperSched}}
\newcommand{\HyperSched}[1]{\textit{HyperSched}}
\newcommand\hs{\textit{HyperSched}}


\newif\ifcomments
\commentstrue
\ifcomments
\newcommand{\joey}[1]{\textbf{\textcolor{magenta}{Joey: #1}}}
\newcommand{\rliaw}[1]{\textbf{\textcolor{red}{Richard: #1}}}
\newcommand{\lisa}[1]{\textbf{\textcolor{blue}{Lisa: #1}}}
\newcommand{\alexey}[1]{\textbf{\textcolor{purple}{AT: #1}}}
\newcommand{\romil}[1]{\textbf{\textcolor{orange}{Romil: #1}}}
\newcommand{\suggest}[1]{\textcolor{violet}{#1}}
\else
\newcommand{\joey}[1]{\textcolor{black}{}}
\newcommand{\rliaw}[1]{\textcolor{black}{}}
\newcommand{\lisa}[1]{\textcolor{black}{}}
\newcommand{\alexey}[1]{\textcolor{black}{}}
\newcommand{\romil}[1]{\textcolor{black}{}}
\newcommand{\suggest}[1]{\textcolor{black}{}}
\fi

%
\renewcommand{\shortauthors}{Liaw et al.}

%
\begin{abstract}

Prior research in resource scheduling for machine learning training workloads
has largely focused on minimizing job completion times. Commonly,
these model training workloads collectively search over a large number of parameter values that control
the learning process in a \textit{hyperparameter search}.
It is preferable to identify and maximally provision the best-performing hyperparameter configuration (trial) to achieve the highest accuracy result as soon as possible.

To optimally trade-off evaluating multiple configurations and training the most promising ones \emph{by a fixed deadline}, we design and build \HyperSched{}---a dynamic application-level resource scheduler to track, identify, and preferentially allocate resources to the best performing trials to maximize accuracy by the deadline.
\HyperSched{} leverages three properties of a hyperparameter search workload overlooked in prior work -- trial disposability, progressively identifiable rankings among different configurations, and space-time constraints -- to outperform standard hyperparameter search algorithms across a variety of benchmarks.

\end{abstract}

\maketitle

\section{Introduction}
\label{sec:intro}

Developing deep learning models is computationally intensive and is quickly becoming a large consumer of compute resources in the cloud. The process of developing a model entails rapid and often parallel exploration and training of multiple alternative model designs, and each design must be at least partially trained to identify the top performers. 
Once a top model design is identified, the model must be fully trained to maximize its performance, which may take days if not weeks \cite{goyal2017accurate}.
As a consequence, it is preferable to devote more resources to training the top candidate model designs earlier in the development process even while exploring alternative designs. 
This prioritization is particularly useful in a setting where a trained model is needed by a certain deadline, such as a nightly release cycle or a conference deadline, as the time constraints necessitate a balance between exploration of model designs and exploitation of the best one.

In this paper, we address the following objective: given a set of model configurations, fixed resources, and a fixed time budget, produce the best-trained
model. We propose \HyperSched{}, an application-level scheduler that is able to dynamically adjust the allocation of available parallel resources between \textit{exploring} model designs and \textit{exploiting} progressively fewer top-performing designs to achieve the highest prediction accuracy within a specified time budget.

\Hypersched{} is deadline-aware and builds upon optimal design exploration work~\cite{asha, HyperBand} to inform resource allocation decisions. \HyperSched{} also addresses several key system level challenges including the scaling limitations of parallel training and the overhead of dynamic resource provisioning. 

We draw upon three key ideas to design our application-level scheduler:

\begin{enumerate}
	\item At the final deadline, only one model configuration is useful, so the results of all other configurations are \emph{disposable}.
	\item As the experiment progresses, promising configurations are \emph{progressively identifiable}.
	\item \emph{Awareness of both resource and time constraints}  enables better resource allocation decisions.
\end{enumerate}

\begin{figure*}[ht]
    \centering
    \begin{subfigure}[h]{0.09\textwidth}
        \centering
        \includegraphics[width=\textwidth]{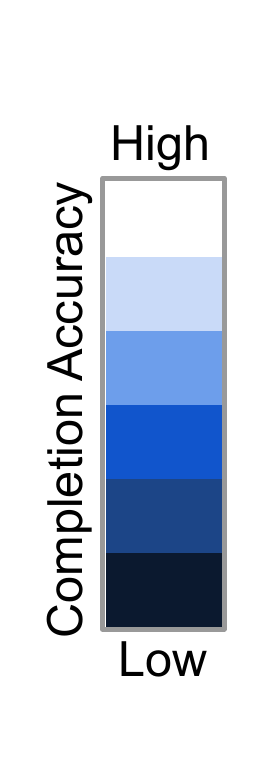}
        \label{fig:spacetime-legend}
    \end{subfigure}
    ~
    \begin{subfigure}[t]{0.265\textwidth}
        %
        \includegraphics[width=0.95\textwidth,right]{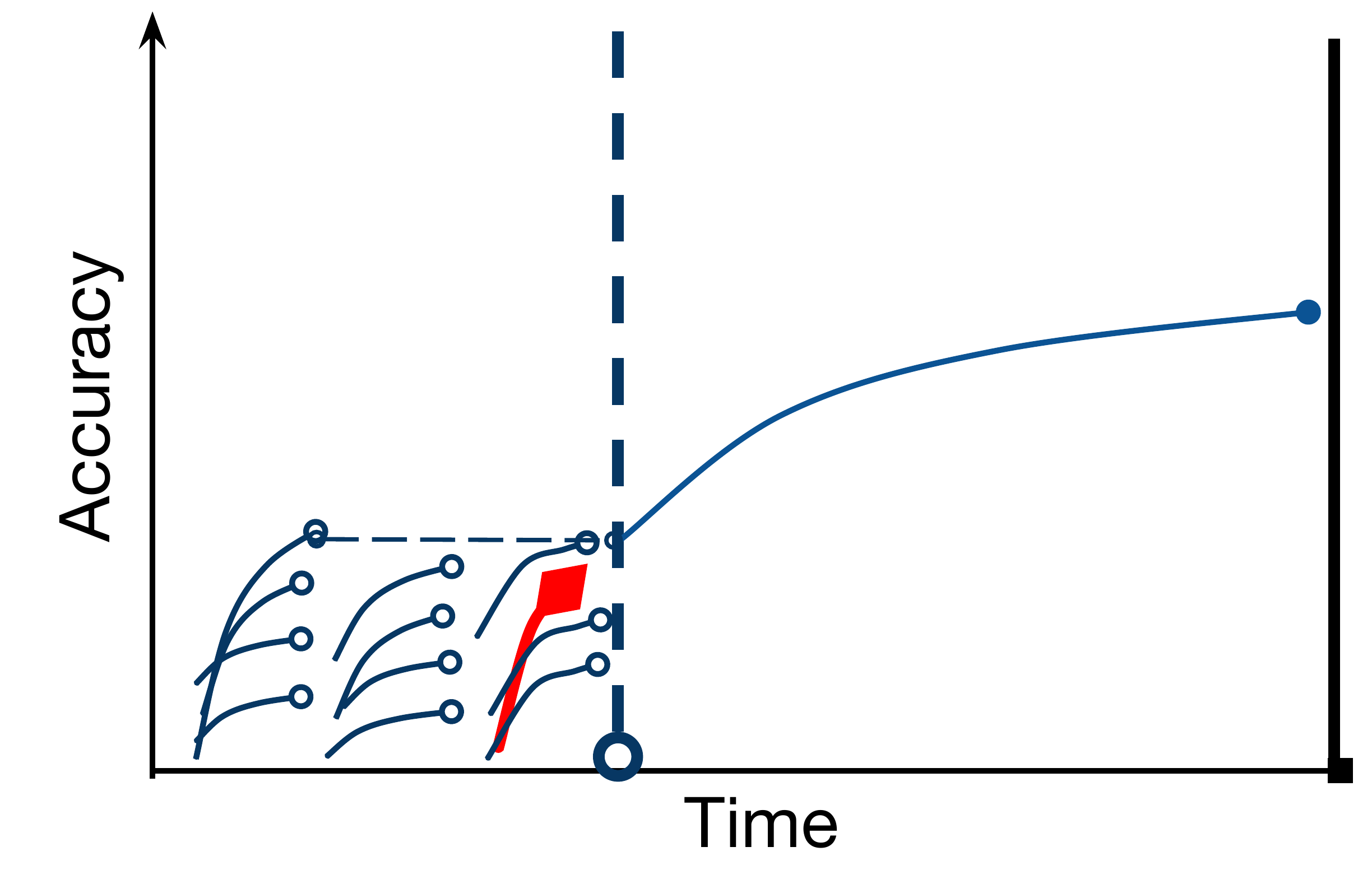}
        \includegraphics[width=\textwidth,right]{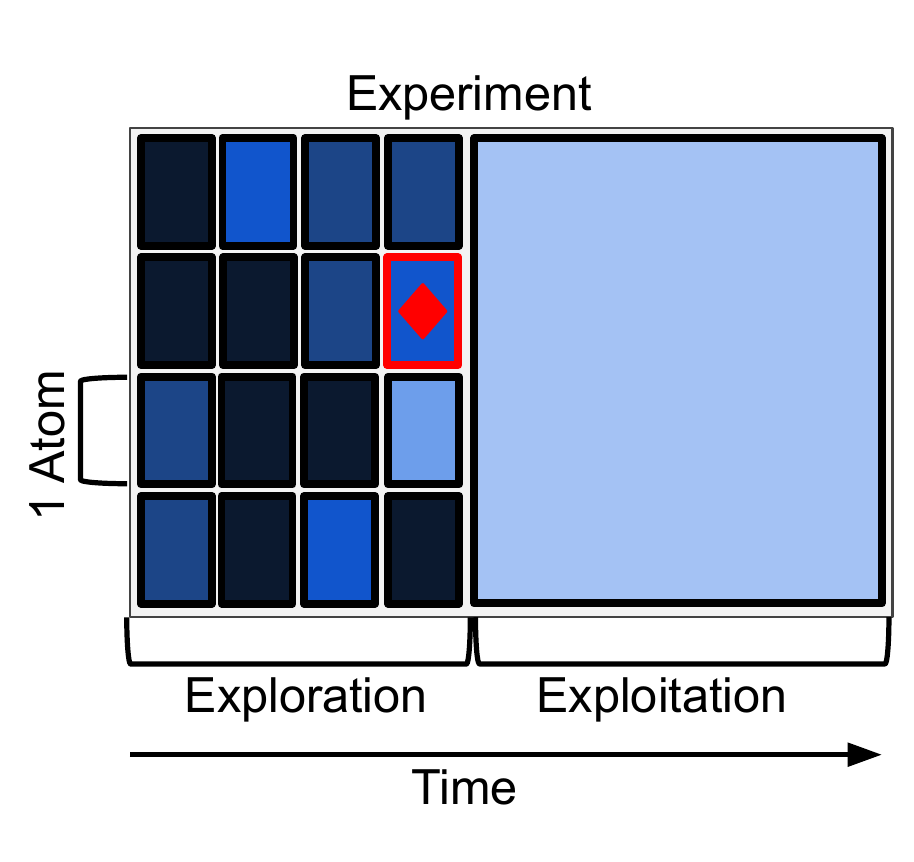}
        \subcaption{Shallow exploitation}
        \label{fig:spacetime-a}
    \end{subfigure}
    ~ 
    \begin{subfigure}[t]{0.28\textwidth}
        \centering
        \includegraphics[width=0.9\textwidth,center]{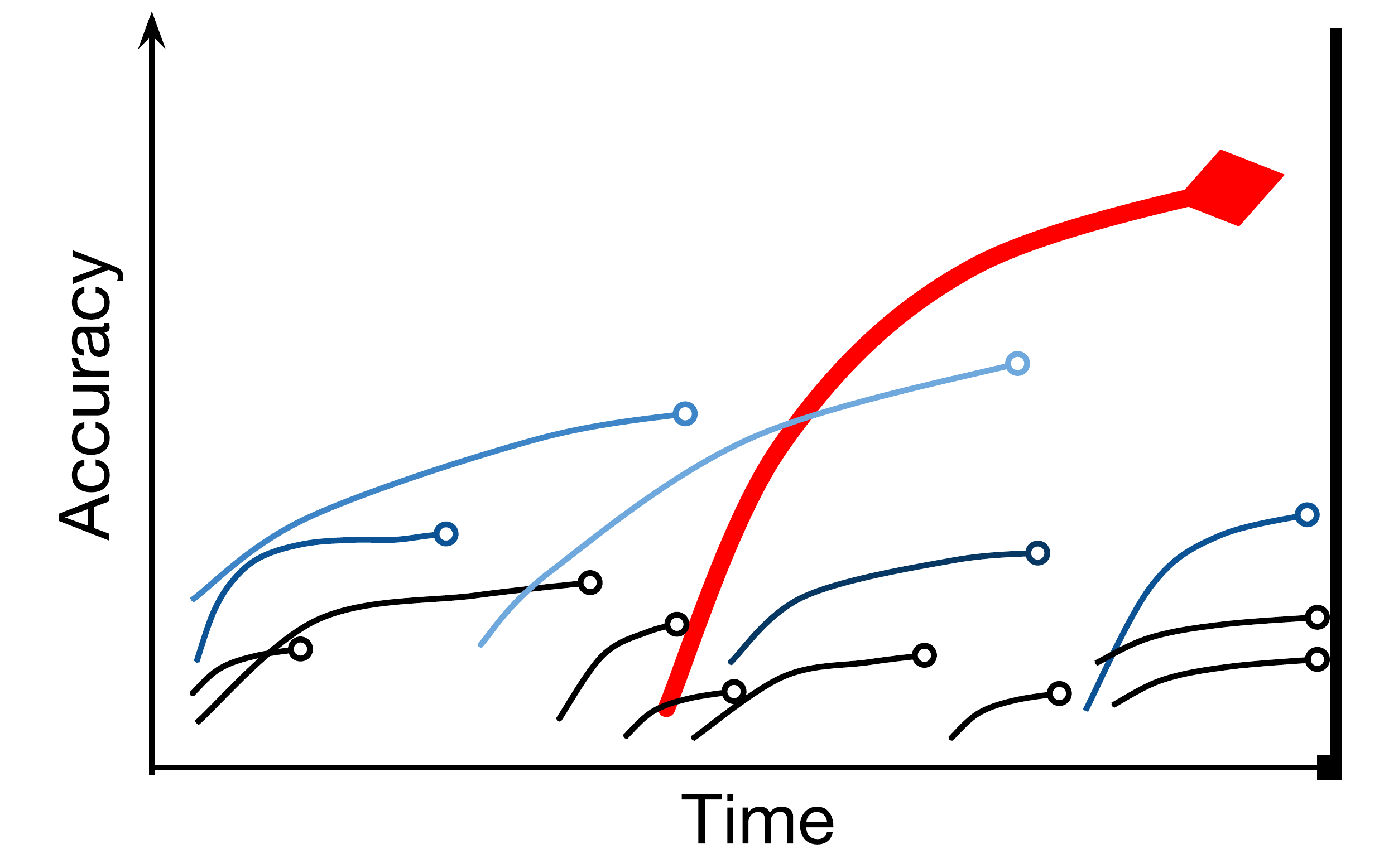}
        \includegraphics[width=\textwidth,right]{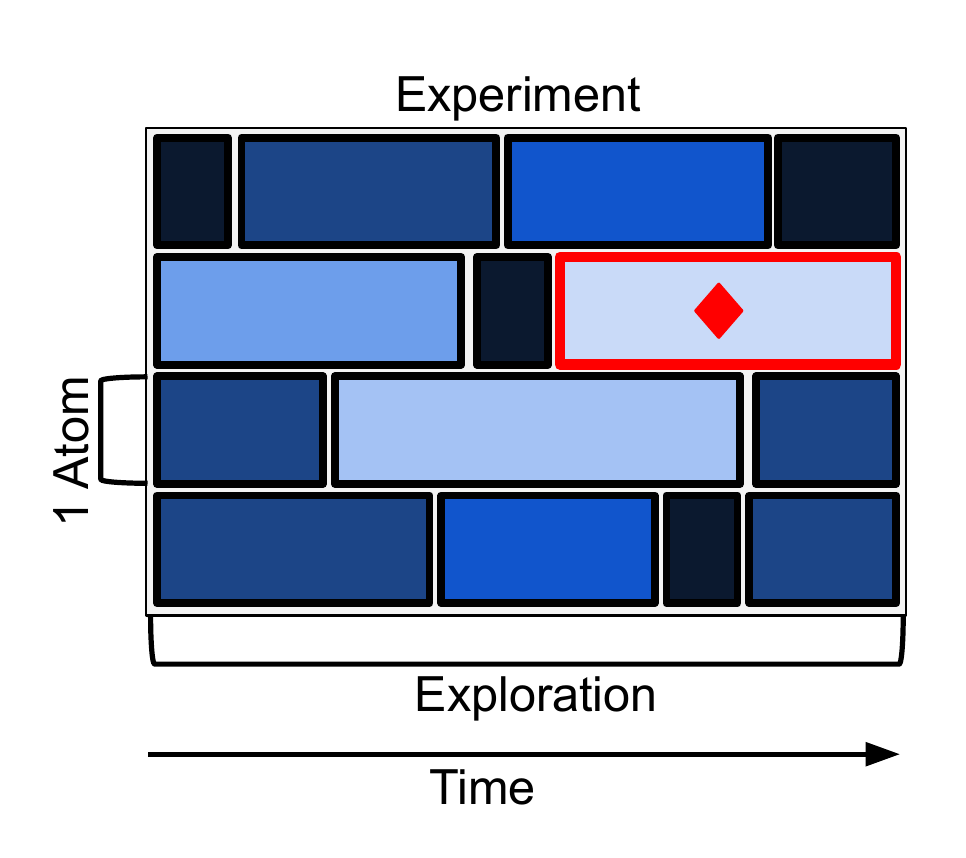}
        \subcaption{Asynchronous Successive Halving}
        \label{fig:spacetime-b}
    \end{subfigure}
    ~ 
    \begin{subfigure}[t]{0.29\textwidth}
        \includegraphics[width=0.9\textwidth,center]{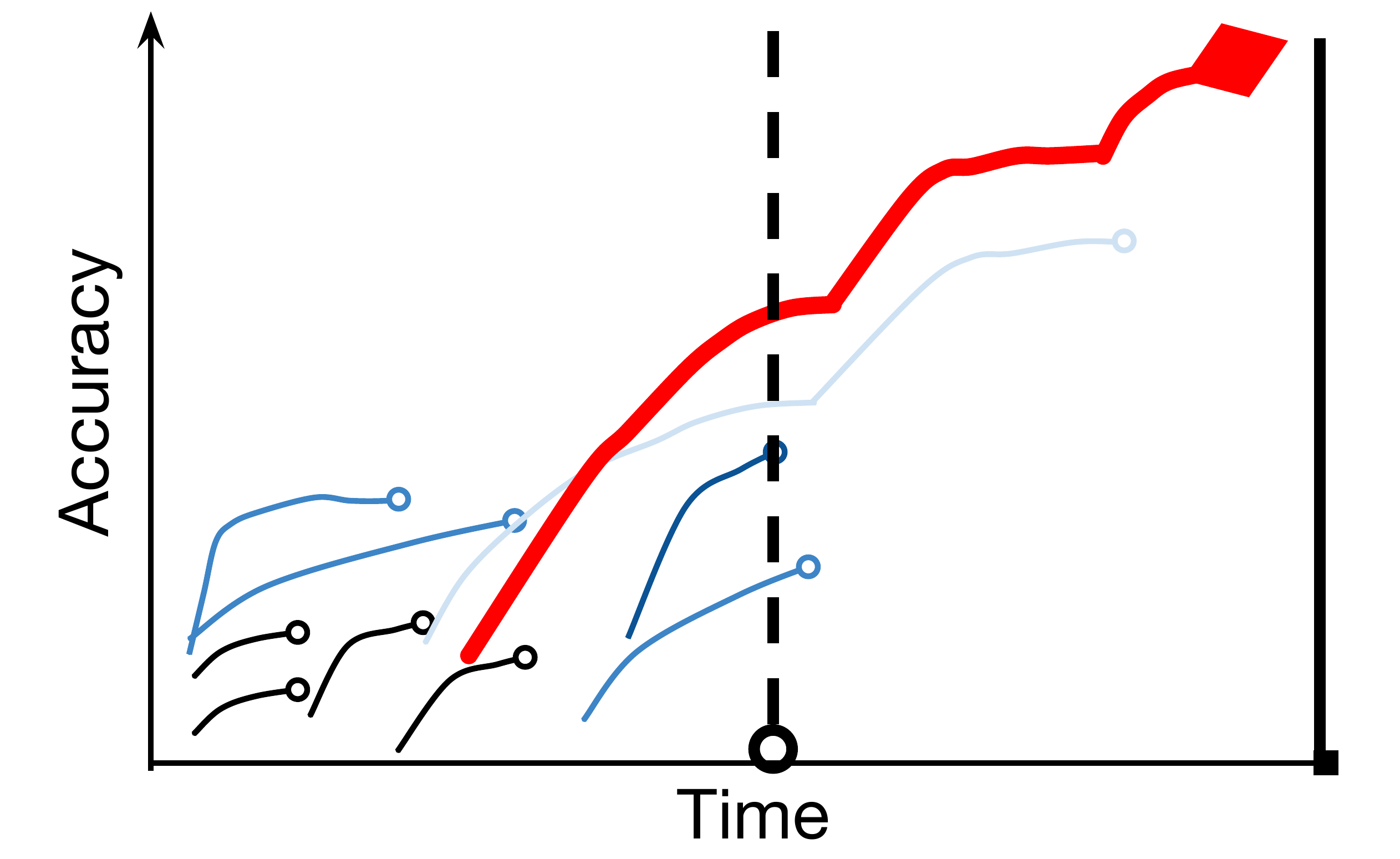}
        \includegraphics[width=\textwidth,right]{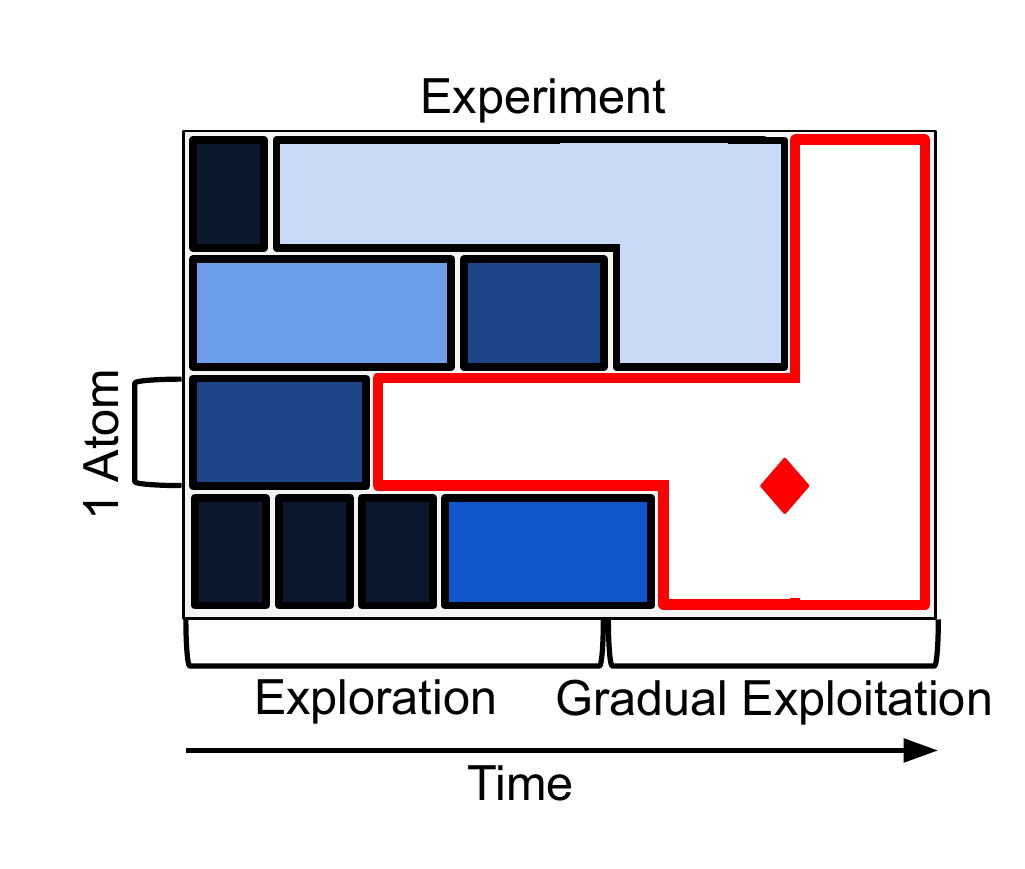}
        \caption{\hypersched{}}
        \label{fig:spacetime-c}
    \end{subfigure}
    \caption{\small A graphical representation of three approaches to resource allocation for hyperparameter search. The above figures for each approach represent the training curves of evaluating multiple trials, and the bottom figures represent the corresponding space-time allocation. The color of the box/trace denotes the accuracy at the end of the trial, where darker indicates lower performing. Atoms are a unit of resource allocation. The trial with the red diamond indicates the same trial, trained for different amounts of time depending on the algorithm. Figure~\ref{fig:spacetime-a} explicitly separates exploration from exploitation. Figure~\ref{fig:spacetime-b} represents a simplified version of a state-of-the-art hyperparameter tuning algorithm (ASHA) that allocates more resource-time to trials that are more promising but is deadline-unaware. 
    Figure~\ref{fig:spacetime-c} represents \HyperSched{}, which is deadline-aware. \HyperSched{} progressively reduces to zero exploration in favor of deeper exploitation of fewer trials by dynamically allocating more parallel resources. }
\label{fig:spacetime}
\end{figure*}




\textbf{Trial Disposability.}
Users will typically run a hyperparameter search (\textit{experiment}), where a sizable number of hyperparameter configurations are evaluated, and one final configuration is chosen, while the other configurations are discarded. We refer to each configuration as a \textit{trial}. This final configuration is often then trained extensively in order to maximize model performance according to a user-specified metric (i.e., model accuracy).

During the hyperparameter search procedure, users or hyperparameter optimization algorithms will use results from individual trials to decide whether to prioritize a certain trial or terminate the execution of a trial.
This disposable nature of trials presents an objective for the application-level scheduler to minimize the amount of ``wasted'' work.

\textbf{Progressively Accurate Identification.} Though each parameter configuration must be at least partially trained to identify the top performers, 
a fraction of the designs can be quickly eliminated without significant training.
Further, for configurations without enough information to classify as promising or poor, we note that there is a generally convergent behavior of deep learning training as more epochs are processed \cite{HyperBand}. 
This presents a dynamic tradeoff space between the cost of exploring more 
jobs against the risk of concentrating all resources in a promising job. 
\HyperSched{} navigates this tradeoff space by leveraging heuristics that indicate a job's performance.

\textbf{Time and Resource Awareness.}
Hyperparameter tuning under time and resource constraints is naturally a constrained optimization problem.
While individual trials can execute faster given more resources, it is also critical that the time and resource constraints are jointly considered, allowing the optimization algorithm to better toggle between exploration, where new trials are evaluated, and exploitation, where promising trials are allocated more resources.
At a high level, by utilizing the knowledge of these constraints, \HyperSched{} can adaptively dedicate more space-time resources to be used for accelerated training of progressively fewer trials, eventually maximizing the accuracy of the best trial.

To summarize, we make the following contributions:
\begin{enumerate}
\item We introduce \Hypersched{}, an application-level scheduler which dynamically allocates resources to trials within an experiment to maximize accuracy at deadline.
\item We implement \Hypersched{} as a scheduler of Tune~\cite{tune}, a distributed hyperparameter tuning framework.
\item We demonstrate significant improvements (up to a 10\% increase in accuracy) under time constraints on various modern deep learning models compared to ASHA~\cite{asha}---a state-of-the-art hyperparameter search algorithm.
\end{enumerate}




\section{Background and Motivation}
Machine learning practitioners are often given time constraints and a finite set of compute resources. Under this space-time resource constraint, machine learning practitioners want to maximize an objective, such as accuracy. Typically, this process requires evaluating large amounts of \textit{hyperparameter} configurations (\textit{trials}) to find a configuration that optimally influences the learning dynamics of the model, followed by extended training of the selected fine-tuned model.

Given these resource constraints and objective, the issue of choosing the right level of parallelism surfaces: how many resources of the given set should be allocated at the level of an individual model, which commonly takes the form of data-parallel training, and how many resources should be allocated to parallel model training in the context of a hyperparameter search? In other words, how do we effectively trade off \textit{exploitation} and \textit{exploration} in the context of time and resource constraints?

\textbf{Parallel training}: Deep learning models often require large amounts of data and take hours if not days to complete training. 
For a single model training run, effectively utilizing multiple parallel compute resources significantly reduces the job completion time.
Deep learning models are most often parallelized in a data-parallel fashion, where replicas of the same neural network model are placed on different devices (i.e., GPUs). Each model computes an update which is then aggregated across all replicas, keeping each replica in sync as iterations progress. Recent research \cite{scaling, jia2018highly} has developed methods for scaling deep learning training jobs to multiple machines with high efficiency by increasing the amount of data fed into generating a single update and simultaneously increasing the magnitude of the update step.

\textbf{Parallel hyperparameter search}: Hyperparameter search aims to identify the single hyperparameter configuration that returns the best optimized objective. The space of these hyperparameters is often very large, and the size of the space is often arbitrarily designated by the user. The process is increasingly important as modern machine learning algorithms such as reinforcement learning algorithms and GANs are highly sensitive to the hyperparameter configuration~\cite{henderson2018deep, goodfellow2014generative}. Random search has not only shown to be highly effective but also easily parallelizable \cite{randomsearch}, enabling users to launch multiple hyperparameter configuration evaluations (\textit{trials}) at once.

\subsection{Exploitation vs Exploration}
When running an experiment testing many different hyperparameter configurations, one can partially train the majority of configurations to determine the best configuration. However, only one configuration, the best, needs to be trained to completion. 

One naive approach to addressing this, as shown in Fig. ~\ref{fig:spacetime-a}, is to spend some of the allocated resource time to try different hyperparameters for a limited duration (explore), pick the best performing, then allocate all resources to that hyperparameter configuration (exploit). 
However, allocating the same short amount of resources to each configuration may not be sufficient to resolve the top configurations. 
Oftentimes, better performing trials at later stages of training may appear similar in performance in initial stages of training \cite{HyperBand}.
In other words, though performance after a small amount of resource allocation does not correspond with absolute performance, there is signal to be gained from comparing relative performance with many alternatives trained with the same amount of resources.

A way to address the need to prioritize trials during exploration is to use the Successive Halving~\cite{successive-halving} algorithm to terminate low performing trials in various stages of training, which allows better performing trials to run longer. Successive Halving
addresses the limitations of Fig. ~\ref{fig:spacetime-a} by introducing a gradation of termination conditions (some are terminated early, some later).


A key problem with the Successive Halving algorithm is that it focuses on identifying the top hyperparameter configuration by a fixed deadline but does not ensure that the best model is fully trained by the deadline. This can be problematic when a developer is running on a fixed deadline and thus must decide how much time to spend finding the best configuration and then train that configuration for the remaining time. 
Because it is deadline-unaware, trials may be started close to the deadline.

In our approach, we address the limitations of these two approaches. Specifically, we extend the Asynchronous Successive Halving Algorithm (ASHA) \cite{asha} by adapting its exploration policy to be deadline-aware and enabling the algorithm to dynamically allocate more resources to well-performing models.



\section{Problem}

\Hypersched{} aims to address one objective: to provide the best-trained
model by a given deadline.
This section will discuss the problem assumptions. We also overview the Successive Halving Algorithm as well as its limitations, which we will later extend to develop \Hypersched{}. Finally, this section will highlight practical issues in implementing dynamic resource allocation.
    
\subsection{Definitions and Assumptions}

A \textit{trial} corresponds to an evaluation of a specific hyperparameter configuration. An \textit{experiment} corresponds to a hyperparameter search composed of multiple trials.
In this work, we focus on scheduling the trials for a single experiment over a fixed set of resources and with a known deadline.
At the deadline, the system must return a model with a good hyperparameter configuration \emph{that has also been adequately trained} so that it can be immediately tested or deployed with maximum accuracy.  

\subsubsection{Assumptions}
Each trial executes an iterative training procedure, and each training procedure will optimize the same objective. 
In an experiment, the configurations of each trial are randomly sampled from the same hyperparameter space. 

Trials are assumed to be \textit{disposable}, meaning that there is no particular trial that must be evaluated to completion or even evaluated in the first place. Cluster fairness is not a design goal, as low performing trials are intentionally starved or eliminated in order to allocate resources more effectively. 

\textbf{Parallelizable training}: Even though \Hypersched{} is able to support a wide range of training workloads, it is targeted towards models that can effectively utilize different quantities of allocated resources and parallelize the training procedure.
Further, we assume that trials themselves may be distributed, and we make the assumption that all trials within an experiment have the same scaling properties. This means that any two trials allocated $N$ resources (i.e., GPUs) will see the same factor of speedup.

\textbf{Framework-specific assumptions:} We assume that trials will return intermediate training progress. Further, model training functions need to support checkpoint-restore functionality, which is commonly offered in many distributed deep learning training frameworks such as TensorFlow and RLlib \cite{tensorflow, rllib}.

\subsection{Successive Halving}

\begin{figure*}[ht]
    \centering
    \includegraphics[width=\textwidth]{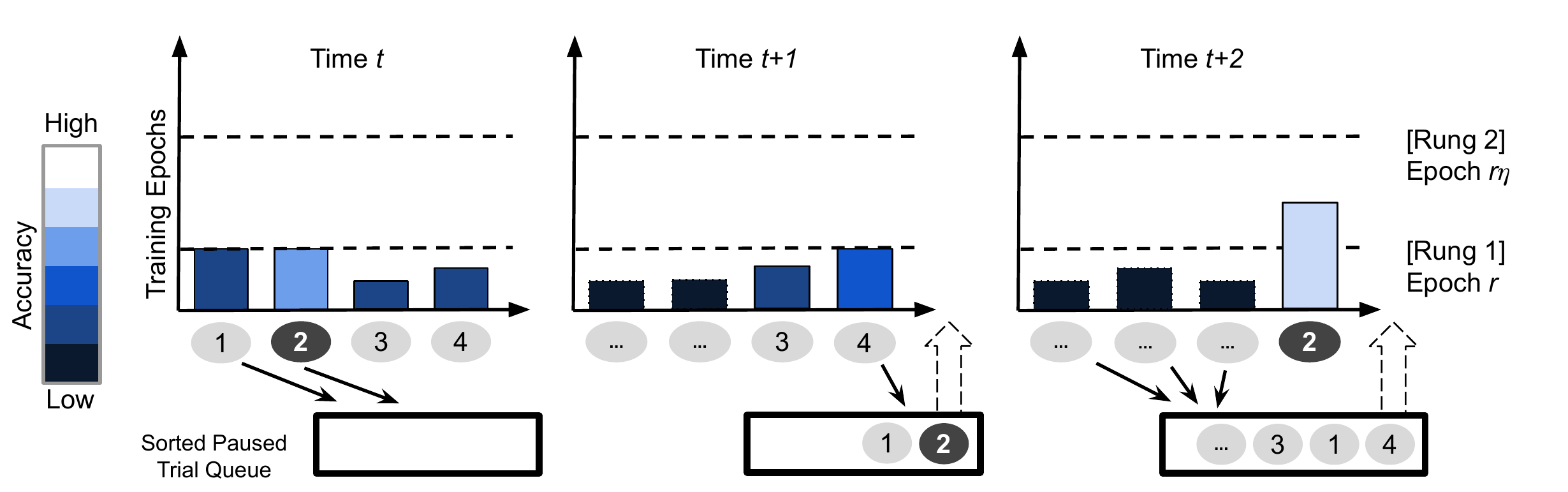}
\caption{\small Illustration of the pausing mechanism in Asynchronous Successive Halving Algorithm (ASHA) with $\eta=3$. At a time step $t$, trials 1 and 2 have been trained for $r$ epochs and are placed in the paused trial queue corresponding to rung 1. The queue is ordered with the better performing trial (trial 2) in the front. At $t+1$, new trials are launched in place of the paused trials, while trial 4 reaches $r$ epochs and is paused. Trial 4 does not perform as well as trial 2. The queue now has enough trials ($\eta$) to promote a trial to the next rung, so trial 2 is unpaused and continued. Later, when the queue has seen $2 * \eta = 6$ trials, it will unpause and continue the next in line. The same process happens at every rung ($r\eta$, $r\eta^2$, etc).}
\label{fig:success}
\end{figure*}
To both identify the most promising configuration and optimize/train it as much as possible by the given deadline, an algorithm needs to balance both exploration and exploitation. For exploration, we draw upon the Successive Halving Algorithm (SHA) \cite{successive-halving}, which underlies recent state-of-the-art bandit-based hyperparameter search techniques, Hyperband and ASHA~\cite{asha} (Asynchronous Successive Halving). 

SHA is designed to address the non-stochastic "best-arm identification problem", where an algorithm is given a fixed budget of "resources" and $N$ "arms" (in our case, trials). Each arm reveals a sequence of scores but requires "resources" to evaluate. The algorithm is then tasked to identify, but \textit{not necessarily exploit}, the "arm" that yields the best result. 
The "resource" value is a theoretical quantity and practically corresponds to resource-time or training epochs.

SHA allocates an increasing number of "resources" to the top fraction of trials in an iterative fashion, termed as a "halving iteration"\footnote{A "halving iteration" roughly correspond to the "rung" concept in ASHA, as described in Section~\ref{sec:asha}.}. The total "resources" allocated per iteration is maintained: 81 trials x 1 "resource" per trial in the first halving iteration, 27 x 3 per trial in the second, 9 x 9, 3 x 27, and finally, 1 trial x 81 "resources" in the last iteration. This allows more promising trials to complete more work in each iteration. The user sets a \emph{reduction factor} that controls the ratio of exploration to exploitation; in the above example of SHA, this reduction factor is $3$.
 
We highlight two practical hyperparameter search algorithms utilizing SHA: Hyperband and ASHA. Hyperband executes SHA in a synchronous fashion, where all jobs are allocated a limited amount of resources first, then the top trials are taken to be executed in the next round of successive halving. Practically, Hyperband's synchronous nature is sensitive to stragglers and dropped jobs, as every trial within a halving iteration must complete consuming its resource before Hyperband can proceed to the next halving iteration.
ASHA provides an asynchronous and parallel algorithm that alleviates this issue. Therefore, for this paper, we will conduct experiments against ASHA. 

\subsection{Asynchronous Successive Halving (ASHA)}
\label{sec:asha}
ASHA takes the following three parameters:
\begin{itemize}
\item $R$, maximum resource allocation (epochs) per trial
\item $r$, minimum resource allocation (epochs) per trial
\item $\eta$, reduction factor
\end{itemize}

For the sake of simplicity, we assume one "resource" corresponds to one epoch in the text. Note that this can be an arbitrary user-defined quantity. Similar to SHA halving iterations, ASHA will let each trial run for at least $r$ epochs, which corresponds to a "rung". At $r$ epochs, ASHA will compare its score to other trial scores after $r$ epochs, and the trial will be paused if it is not in the top $1/\eta$ of evaluated scores. The same evaluation procedure will take place at $r\eta$ epochs, $r\eta^2$, ... $r\eta^i$ - each corresponding to a different "rung". After $R$ epochs, the trial will terminate. The reduction factor $\eta$ controls the ratio of exploration to exploitation. 

When a trial is paused, the scheduler becomes underfilled. ASHA will fill the cluster by resuming a paused trials if its score is in the top $1/\eta$ of evaluated scores in its rung. Rungs will be checked in descending order. If no trial is found, ASHA will evaluate a new trial, thereby exploring new trials throughout its lifetime while prioritizing exploitation of trials in the higher rungs. A graphical illustration of the pausing mechanism is shown in Figure~\ref{fig:success}.

\subsubsection{Limitations in the deadline setting.}
ASHA, being deadline-unaware, will continue starting new trials through the execution of the job. The likelihood of any of these new trials improving the objective at deadline decreases as the deadline draws closer. Further, randomly selecting a trial that will do better than the current best trial becomes increasingly rare as more trials are evaluated. 

Another limitation of ASHA is its pausing criteria. ASHA maintains a $1/\eta$ ratio of continued trials and will pause trials if they are within the first $\eta$ trials to reach a rung. However, in the time-constrained setting, this invariant can clog the exploration process, resulting in degraded performance.

Finally, because the Successive Halving algorithm is focused on best-arm identification, it does not specifically allocate time for the actual exploitation of the identified best-arm. The diminishing value of exploring new trials motivates an opportunity to improve ASHA by adaptively decreasing the number of trials seen in favor of exploiting currently running trials.




\subsection{Dynamic Resource Allocation}
\label{sec:dra-problems}

As mentioned above as a specific limitation of ASHA, it is possible to accelerate specific trials by dynamically allocating more resources to them. However, prior work \cite{scaling} has empirically demonstrated the inability to effectively \textit{and} robustly utilize these resources to accelerate training due to two constraints.
First, network and communication overheads can dominate at certain scales, so scaling does not necessarily increase linearly. Second, checkpoint and restore overheads of deep learning models can be significant. Especially under deadline constraints, an allocation policy needs to account for overheads and effective scaling to improve performance. Thus, any dynamic resource reallocation policy must be \textit{scale-aware} and \textit{overhead-aware}.



\section{HyperSched}

In this section, we discuss the design of \hs{}. At a high level, \hs{} extends ASHA for constrained resource space-time setting by using an exploration policy that is deadline-aware and by dynamically allocating more resources to fewer promising training trials.

\hypersched{} requires the user to specify a logical unit of resource allocation, or \textit{atom}, along with a total available count of atoms $N$. \hypersched{} reallocates resources in integer quantities of atoms, up to $N$. For example, users can specify 1 GPU as a \textit{atom} for distributed deep learning training, and \hypersched{} also supports other quantities such as fixed CPU and GPU pairings. 

\subsection{Overview of \hs{}}
The inputs to \hs{} are similar to ASHA but require three more parameters from the user - the time deadline, total resource atoms available, and a scaling function $s(a)$. $s(a)$ is a function that maps atoms to epochs per second (or any rate of progress) and does not need to be exact. 

\begin{itemize}
\item Deadline $T$
\item Resource atoms $N$
\item Scaling function $s$
\item Minimum resource allocation $r$ (same as ASHA)
\item Maximum resource allocation $R$ (same as ASHA)
\item Reduction factor $\eta$ (same as ASHA)
\end{itemize}

Note that these parameters are not hyperparameters needed for tuning but simply job parameters that are natural of many job allocations.

Similarly to ASHA, a trial that \hypersched{} evaluates will run for at least $r$ epochs, and its score will be compared to other trials at $r$, $r\eta$ , $r\eta^2$, ... (at each rung). At every evaluation point, trials will either continue executing or will be paused. Trials will be terminated after $R$ epochs. The reduction factor $\eta$ controls the ratio of exploration to exploitation.

When a trial is paused, the cluster becomes under-filled (i.e., the number of resource atoms used is less than $N$), and \hypersched{} will either: 
\begin{enumerate}
    \item try to find and execute a paused trial in the top $1/\eta$ of a rung (similar to ASHA), where rungs are checked in descending order
    \item generate a new trial by sampling from the hyperparameter search space (similar to ASHA)
    \item allocate more atoms (resources) to an existing trial
\end{enumerate}

\begin{figure}[t]
  \includegraphics[width=\columnwidth]{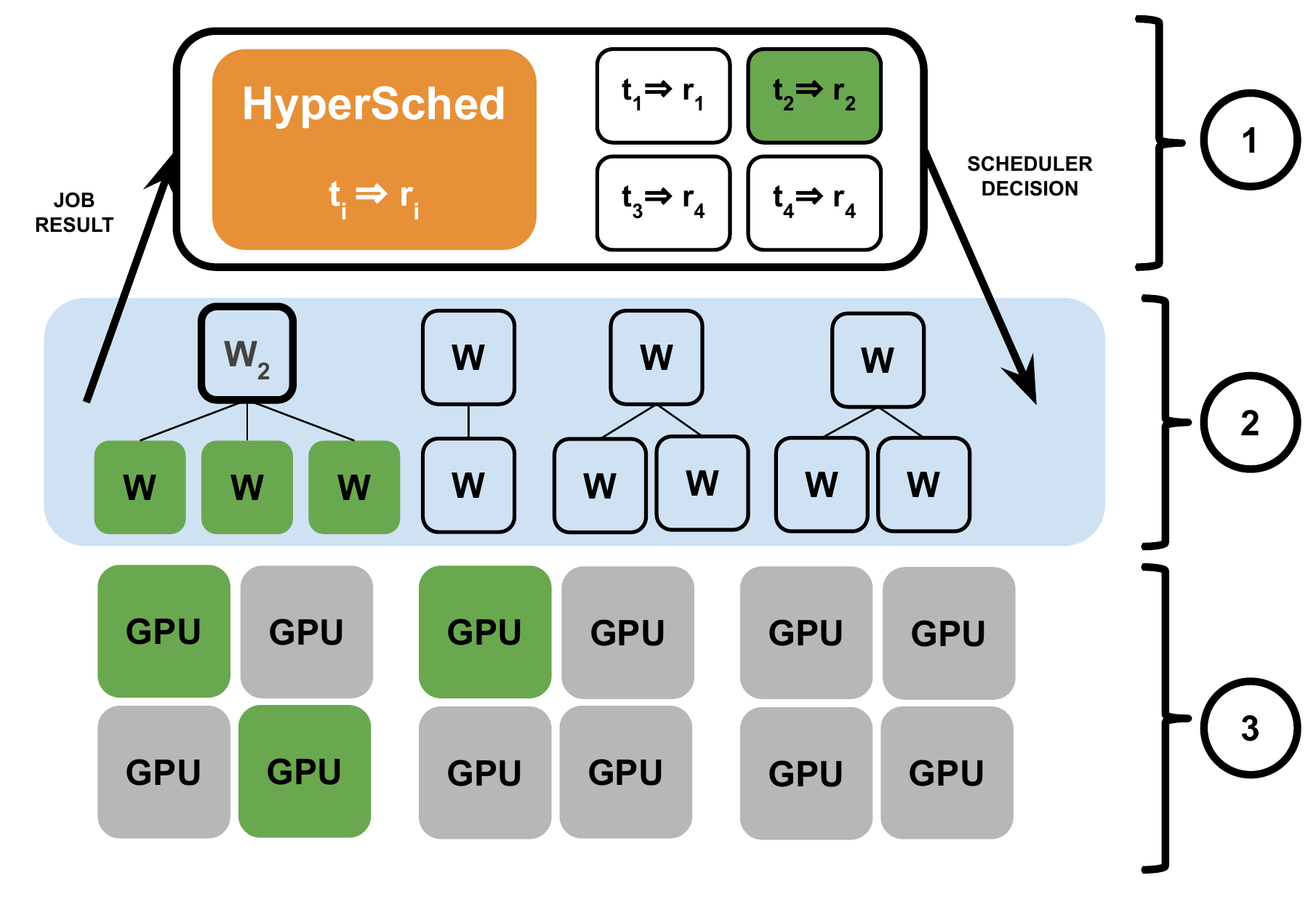}
  \captionsetup{belowskip=0pt}
  \caption{\small \hypersched{} system architecture. \hypersched{}, implemented in the Tune framework (1), maintains a mapping $t_i$ to $r_i$, along with trial metadata. Each trial has a logical representation which receives a resource allocation. Each trial has one main coordinating actor ($w_i$), which may delegate its resource allocation to sub-actors. The execution of each actor and trial is done with Ray (2) and placed on physical devices (3). The specific placement is not handled by \hypersched{}. The intermediate results are retrieved from actors and passed to \hypersched{} (1), which uses those results to update its resources $r_i$, and then provide a decision for Tune to execute. }
\label{fig:systemarchitecture}
\end{figure}

Below, we will discuss the mechanisms in \hypersched{} that decides between these choices.

\subsection{Exploration}
Given finite resource and time constraints, it is necessary to be \textit{deadline-aware} for a more effective trade-off of exploration and exploitation. For this, \hypersched{} utilizes \textit{speculative evaluation} and a \textit{deadline-aware entrance policy}.

\textbf{Speculative evaluation:} At each evaluation point/rung, \hypersched{} will pause a trial if it is not in the top $1/\eta$ of all currently \textbf{seen} trials of that rung. This means that unlike ASHA, a rung's first-arrived trial will not be paused, despite the rung having not seen any previous trials. Note that by the above modification, it is possible for the first-arrived trials to be low-performing and result in blocking the evaluation of other, possibly better-performing configurations. To avoid such head-of-line blocking, \hypersched{} will continuously monitor and compare the performance of previously-arrived trials to new trials. If at any time the trial score at the rung is not in the top 1$/\eta$ of all \textbf{seen} trials at the rung, the trial will be paused.



\textbf{\hypersched{} Entrance Policy:} 
\hypersched{} will stop running new configurations after a certain threshold. Trials that are launched too close to the deadline will have a negligible probability of changing the maximum score by the deadline, and more resources can be dedicated to currently running jobs for a higher chance of increasing the highest accuracy. Further, in a random search, the marginal improvement of maximum accuracy diminishes as the number of trials evaluated increases.

Specifically, \hypersched{} will track the longest duration of each running trial $T'$, the time left until the deadline $T_n$  and the time per epoch using one atom ($T_a$). If $\min{\{R*T_a, \eta T'\}} < T_n$, which is the minimum between the time it takes to run one new trial to max epochs $R$ and $\eta$ times the longest running trial duration is less than time left till deadline, \hypersched{} will not continue to launch new trials. This is done in the same spirit as SHA. In SHA, an exponentially increasing distance for identifying the ordering between two configurations is theoretically justified.

\subsection{Exploitation}
\label{sec:hs-exploit}

\begin{figure}[t]
    \centering
    \begin{subfigure}[t]{0.23\textwidth}
        \centering
        \includegraphics[width=\textwidth]{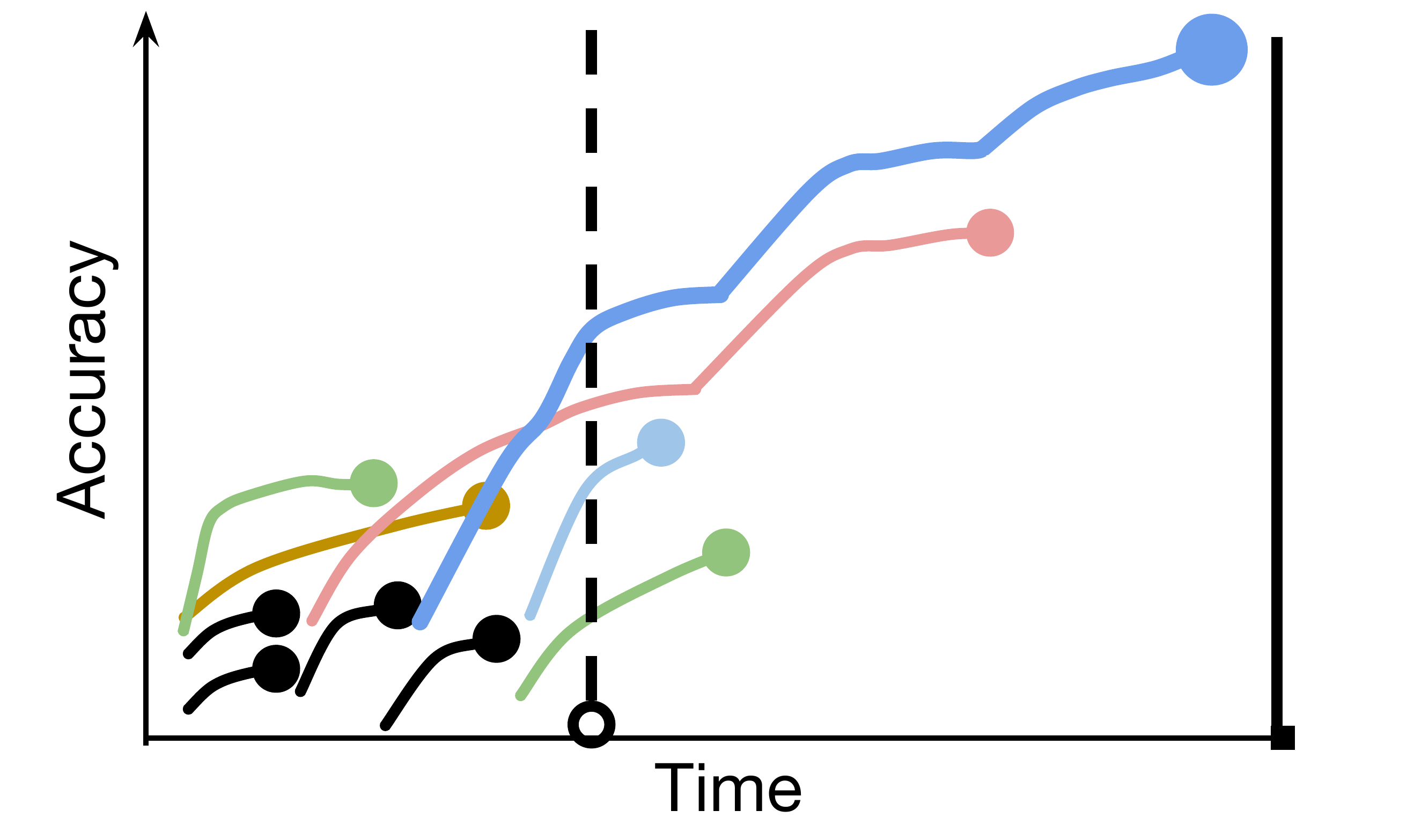}
        \captionsetup{justification=centering}
        \caption{Graphical representation of trial accuracy over time. }
        \label{fig:hs-curve}
    \end{subfigure}
    \begin{subfigure}[t]{0.23\textwidth}
        \centering
        \includegraphics[width=\textwidth]{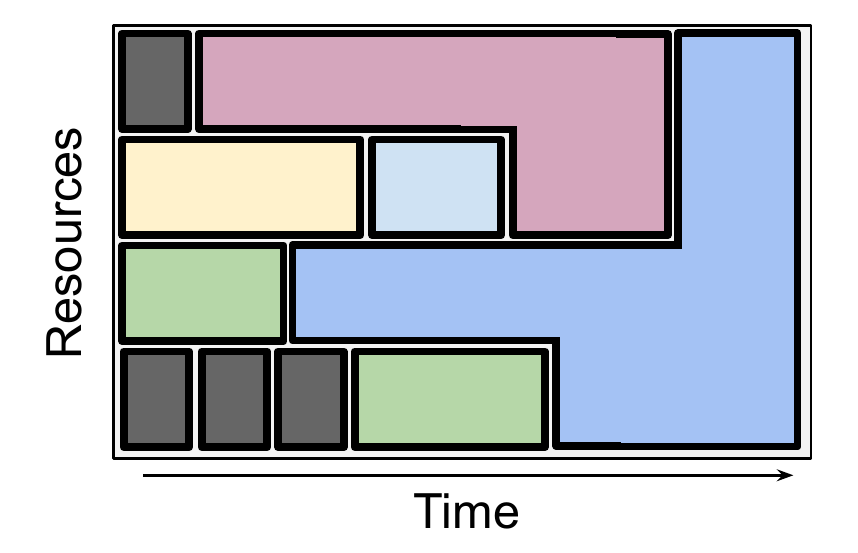}
        
        \captionsetup{justification=centering}
        \caption{Corresponding space-time allocation of trials.}
        \label{fig:hs-space}
    \end{subfigure}
  \caption{\small Accuracy and space-time representation of the exploration/exploitation properties of \hypersched{}, which progressively reduces/eventually eliminates exploration in favor of deeper exploitation of increasingly fewer top trials. Trial accuracy in Figure~\ref{fig:hs-curve} (higher is better) is color-mapped to the corresponding resource allocation. In Figure~\ref{fig:hs-space}, the height of the boxes represents resource allocation to that particular job, showing the increasing number of resources allocated to increasingly fewer top performing trials. Dotted line in Figure~\ref{fig:hs-curve} indicates when new trial exploration stops, determined by the deadline.}
  \label{fig:hypersched-spacetime}
\end{figure}

Unlike SHA, which aims to address the problem of best arm \textit{identification}, \hypersched{} aims to optimize the highest accuracy at deadline, or "best configuration \textit{exploitation}". 
This is done via a exploration/exploitation policy that incorperates \textit{dynamic resource allocation} as illustrated in Fig. ~\ref{fig:hypersched-spacetime}.

\textbf{Dynamic Resource Allocation}: When the cluster is underutilized (i.e, there are more atoms than trials left), atoms are uniformly allocated across the currently running trials. We choose a uniform allocation policy for consistency with the theory behind best-arm identification that is used by in successive halving. However, we note that depending on the distribution of the workload performance, a more exploitative policy such as one that allocates all resources to the top trial may be more effective.

\textbf{Accounting for Overheads:} As mentioned in Section~\ref{sec:dra-problems}, overheads of the resizing may be significant. To address this, users can profile scaling before beginning the search by sampling different allocations on a random trial, while monitoring the startup time of each trial. Before executing the resize, \hypersched{} checks that the proposed new allocation would increase the total amount of samples. Namely, let $T_o$ be the startup time (in seconds), $T$ is the time left until deadline, and $H$, $H'$ is the throughput (i.e., epochs per second) of the model training procedure given the current allocation and the proposed allocation. Then, \hypersched{} will execute the resizing if $H'(T-T_o) > HT$.

    

\section{System Implementation}

\IncMargin{0.5em}
\begin{algorithm}
\small
\SetKwFunction{experiment}{runExperiment}
\SetKwProg{algo}{Algorithm}{}{}
\SetKwProg{func}{Function}{}{}
\SetKwFunction{startnexttrial}{start\_next\_trial}
\SetKw{Continue}{continue}
\SetKw{break}{break}
\SetKw{and}{AND}
\textbf{Input: } {scheduler, deadline $T$}\\

\algo{\experiment{scheduler}} {
    $liveTrials$ = [run\_new\_trial()]\;
    \While{$time$.now() < $T$}{
        \If{$liveTrials$.empty()}{\break}
        \If{{$scheduler$.entrance\_policy($liveTrials$)}}{
            $start\_next\_trial(scheduler.rungs)$\;
            \Continue
        }
        $trial$ = get\_available\_trial($liveTrials$)\;
        $decision$ = $scheduler$.schedule($trial$)\;
        \eIf{$decision$ in [PAUSE, STOP]} {
            $trial$.set\_state($decision$)\;
            $liveTrials$.remove($trial$)\;
        }{$trial$.run()\;}
    }
    \Return $trial$ with highest accuracy
}

\func{\startnexttrial{rungs}}{
    \For{rung in descending\_sort(rungs)}{
        top\_k = floor(len(rung.trials) / $\eta$)\;
        \For{trial in sorted(rung.trials)[0:top\_k]}{
            \If{trial.paused} { 
                $trial$.run()\;
                \Return $trial$
            }
        }
    }
    \Return run\_new\_trial()
}

\caption{Experiment execution loop in Tune}
  \label{alg:run_experiment}
\end{algorithm}
\DecMargin{0.5em}  

We have implemented \hs{} as a component of Tune \cite{tune}, an open-source distributed hyperparameter search framework. \hypersched{}'s integration with this framework is illustrated in Fig. ~\ref{fig:systemarchitecture}. Tune leverages Ray, a distributed execution engine, launching parallel training jobs using Ray's actor API. We relegate the physical placement on specific devices of trials (and their subworkers) to Ray, which has its own lower-level task placement and scheduling mechanisms.

Tune takes in a hyperparameter space and a number of trials to run and execute. Each trial executes on a separate Python process separate from the Tune process and will execute a user-designated quantity of training (i.e., 1 epoch). Trials themselves can be parallel or distributed, possibly starting their own set of processes for parallel training.

Tune provides a custom scheduler interface, which can make decisions based on trial performance. At the end of each training step, the scheduler will be notified and will affect the training process by choosing from a set of primitives (\texttt{pause}, \texttt{stop}, \texttt{continue}). 

Algorithm~\ref{alg:run_experiment} provides an overview of the Tune framework to run a hyperparameter tuning experiment. In this framework, trials are populated based on a \textit{entrance policy} specified by the scheduler (ASHA or \hs{}).  Following this, the scheduler picks a trial to allocate resources and run. \texttt{trial.run()} executes one epoch/iteration of training. At the completion of \texttt{trial.run()} , the scheduler is invoked to evaluate if the trial should be stopped, paused or continued.

\textbf{ASHA Implementation:} Algorithm~\ref{alg:asha} describes ASHA as a pluggable scheduler in Tune. ASHA employs a simple entrance policy which will return True when the cluster is underfilled. ASHA runs each trial and pauses it if it is at a rung (\texttt{rung.iter}) and is not in the top percentile of trials, as described in Section~\ref{sec:asha}. Pausing allows either new trials to be admitted or paused trials to continue to be evaluated. This process of suspend-resume is repeated until trials complete max epochs $R$. Note that because ASHA is deadline-unaware, it allows trials to be started very close to the deadline.

\IncMargin{0.5em}
\begin{algorithm}
\small
\SetKwFunction{asha}{ASHA.schedule}
\SetKwFunction{entrancePolicy}{ASHA.entrance\_policy}
\SetKwProg{algo}{Algorithm}{}{}
\SetKwProg{func}{Function}{}{}
\SetKw{Continue}{continue}
\textbf{Input: } {reduction factor $\eta$, total atoms $N$, max epochs $R$}\\
\func{\entrancePolicy{liveTrials}}{
    \tcc{Trivial entrance policy}
    \Return len(liveTrials) < N
}
\func{\asha{trial}} {
    \If{trial.iter > R} {
        \Return STOP
    }
    \For{rung in self.rungs}{
        \If{trial.iter == rung.iter}{
            rung.record(trial.currentAcc)\;
            top\_k = floor(len(rung.trials) / $\eta$)\;
            \If{trial not in rung.trials[0:top\_k]} {
                \Return PAUSE\
            } 
        }
    \Return CONTINUE\;
    }
}
\caption{ASHA with Entrance Policy. Rungs are implicitly instantiated and depend on $r$ and $\eta$.}
  \label{alg:asha}
\end{algorithm}
\DecMargin{0.5em}

\textbf{\HyperSched{} Implementation:} Algorithm~\ref{alg:hypersched2} describes the implementation of \hs{}. The entrance policy of \hs{} is both resource-aware and deadline-aware. The policy checks if there are sufficient available resources in the cluster to admit new trials and also ensures that admitting new trials is beneficial given the remaining time.
Unlike ASHA, \hs{} performs \textit{speculative evaluation} by allowing first-comers to a rung to continue execution. Finally, \hs{} dynamically increases the resource allocation for the most promising trials by estimating the gain from doing so.

During execution, \hypersched{}  maintains a mapping of trials to atoms. This is updated with intermediate trial performance during training. As mentioned in Section~\ref{sec:hs-exploit}, resource allocation per trial will grow once the cluster becomes under-filled. Practically, the uniform allocation is implemented by sorting the running trials by performance and then allocating one atom to each trial in the sorted list in a round-robin fashion. As a result, the top trials will be the first to receive leftover atoms.

When a job is eligible for reallocation, \hypersched{} will first checkpoint the current state of the model, terminate all of the processes for that specific training job, relaunch the training job with the new allocated number of resources and restore the training job to the previous checkpoint. To prevent reallocating resources to one job and incurring large amounts of overhead costs without any progress, \hypersched{} maintains an internal timer for each job that must pass a threshold (\texttt{COOLDOWN})  before resources for the job can be reallocated. During training, \hypersched{} also tracks the cost to start and resize a trial in \texttt{update\_timings}. This median cost is used in allocation decisions as $T_o$ to ensure that the trial is not worse off after the reallocation decision is made.





\IncMargin{0.5em}
\begin{algorithm}
\small
\SetKwFunction{hyperschedfunc}{HS.schedule}
\SetKwFunction{entrancePolicy}{HS.entrance\_policy}
\SetKwFunction{uniformAllocation}{uniformAllocation}
\SetKw{and}{AND}
\SetKwProg{algo}{Algorithm}{}{}
\SetKwProg{func}{Function}{}{}

\SetKw{Continue}{continue}

\textbf{Input: } {reduction factor $\eta$, total atoms $N$, max epochs $R$, deadline $T$, scaling $s(atoms)$ }\\
$T_n = T - time.now()$ \\
$T_o$ = measured resizing overhead \;
$T_a$ = measured time for one epoch with one atom\;
\func{\entrancePolicy{liveTrials}}{
    allocated = $\sum\limits_{i} liveTrials[i].atoms$\;
    $t_f$ = furthestTrial().time\;
    should\_enter = $min(R*T_a, t_f * \eta) < T_n$\;
    \Return allocated $< N$ \and should\_enter
}

\func{\hyperschedfunc{trial}} {
    update\_timings($trial$)\; 
    \If{trial.iter > R} {
        \Return STOP
    }
    \For{rung in self.rungs}{
        \If{trial.iter == rung.iter}{
            rung.record(trial.currentAcc)\;
        }
        \If{trial.iter >= rung.iter}{
            cutoff = top 1/$\eta$ of rung.scores\;
            \If{trial.accuracy[rung.iter] < cutoff} {
                \Return PAUSE
            } 
        }
    }
    atoms$'$ = uniformAllocation($liveTrials$, $N$)[$trial$]\; 
    $expected\_epochs$ = $(T_n - T_o) * s(atoms')$ \;
    \If{expected\_epochs >$T_n * s(trial.atoms)$}{
        \If{trial.iters\_since\_resize > COOLDOWN}{
            $trial$.resize(atoms$'$)\;
        }
    }
    \Return CONTINUE
}
\caption{\HyperSched{} with Entrance Policy. Rungs are implicitly instantiated and depend on $r$ and $\eta$.}
  \label{alg:hypersched2}
\end{algorithm}
\DecMargin{0.5em}

%
\section{Evaluation}

\begin{figure*}[ht]
    \begin{subfigure}{0.24\textwidth}
        \includegraphics[width=1\linewidth] {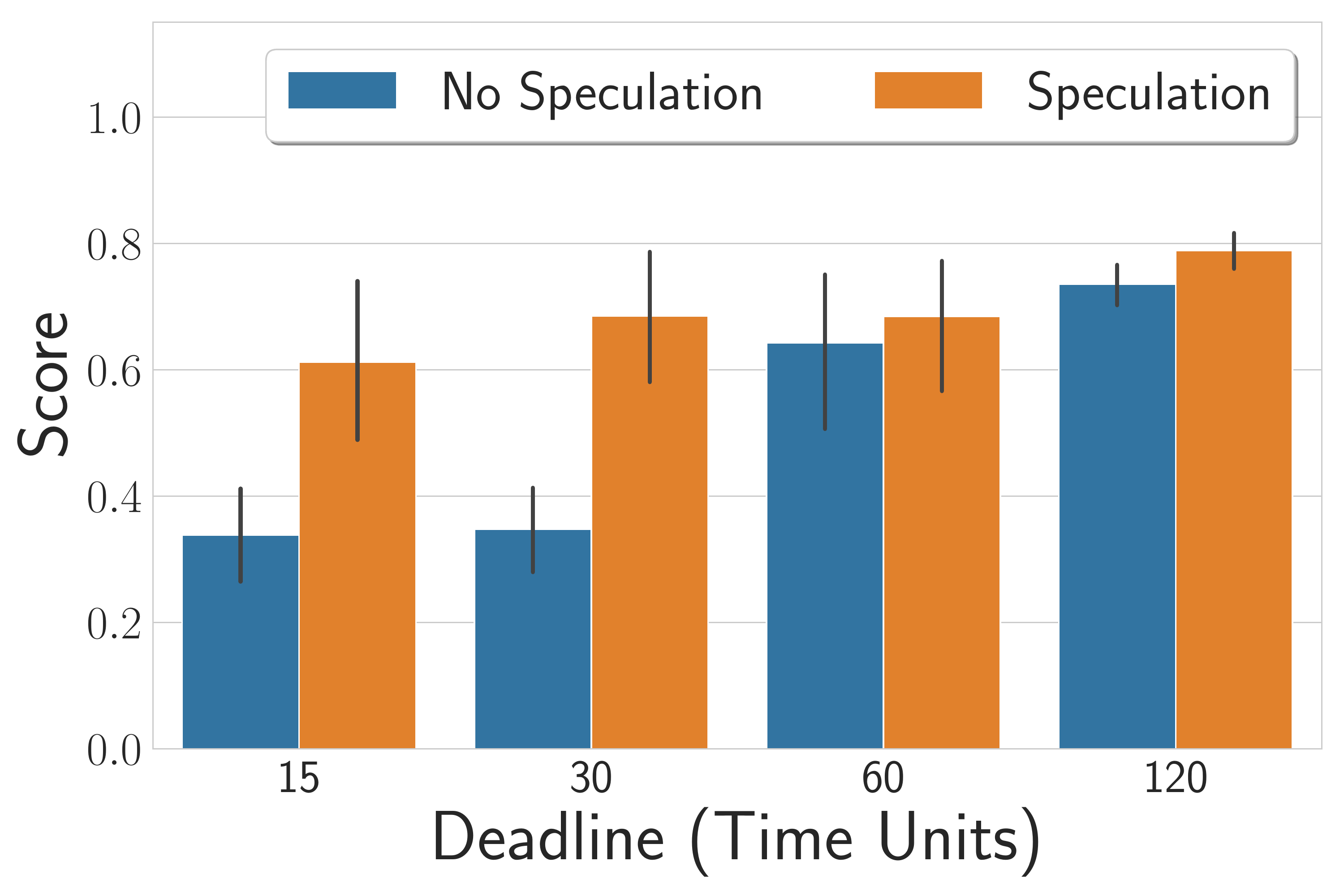}
        \caption{Trial count (4 atoms)}
    \end{subfigure}
    \hfill 
    \begin{subfigure}{0.24\textwidth}
        \includegraphics[width=1\linewidth] {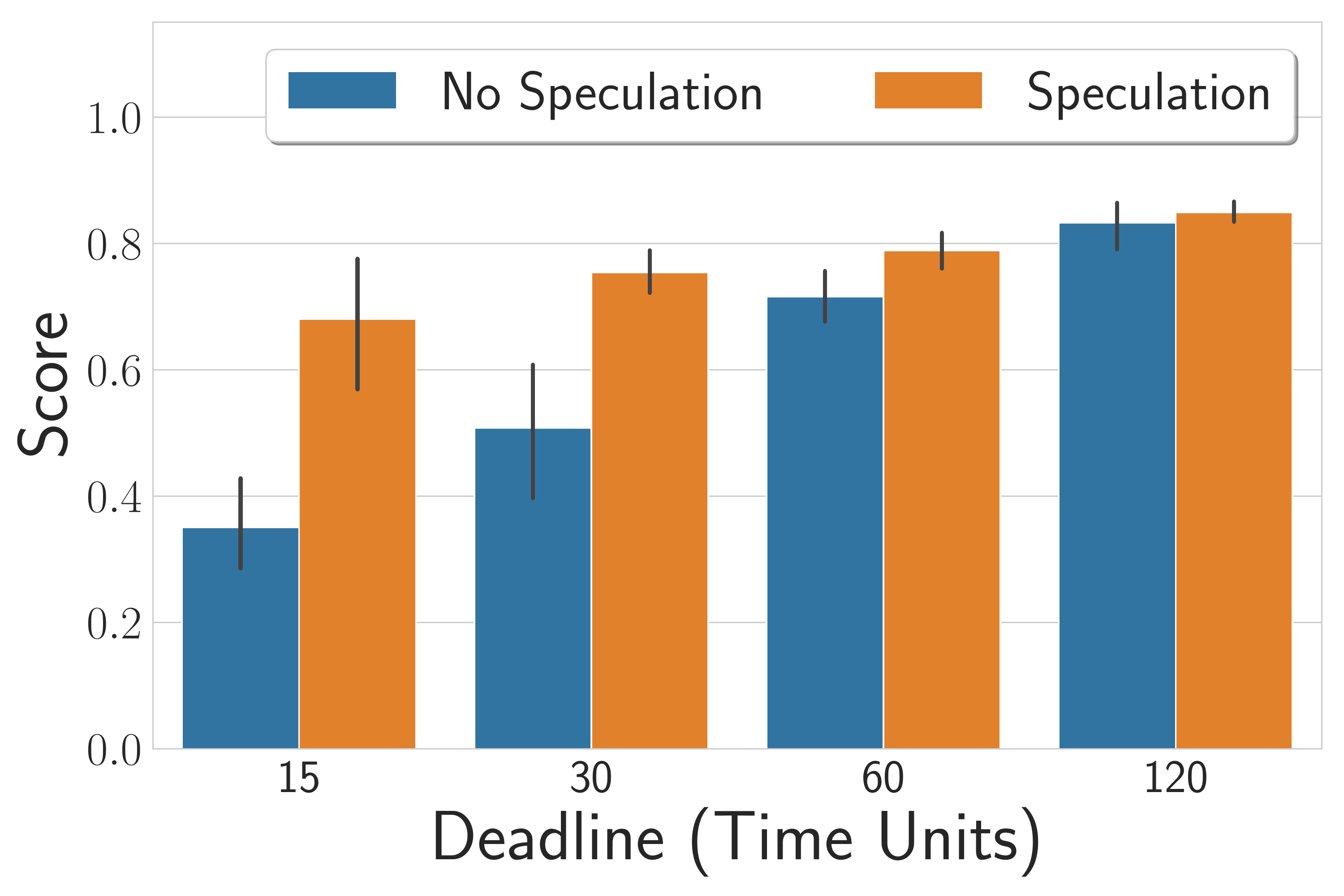}%
        \caption{Trial count (8 atoms)}
    \end{subfigure}
    \begin{subfigure}{0.24\textwidth}
        \includegraphics[width=1\linewidth]{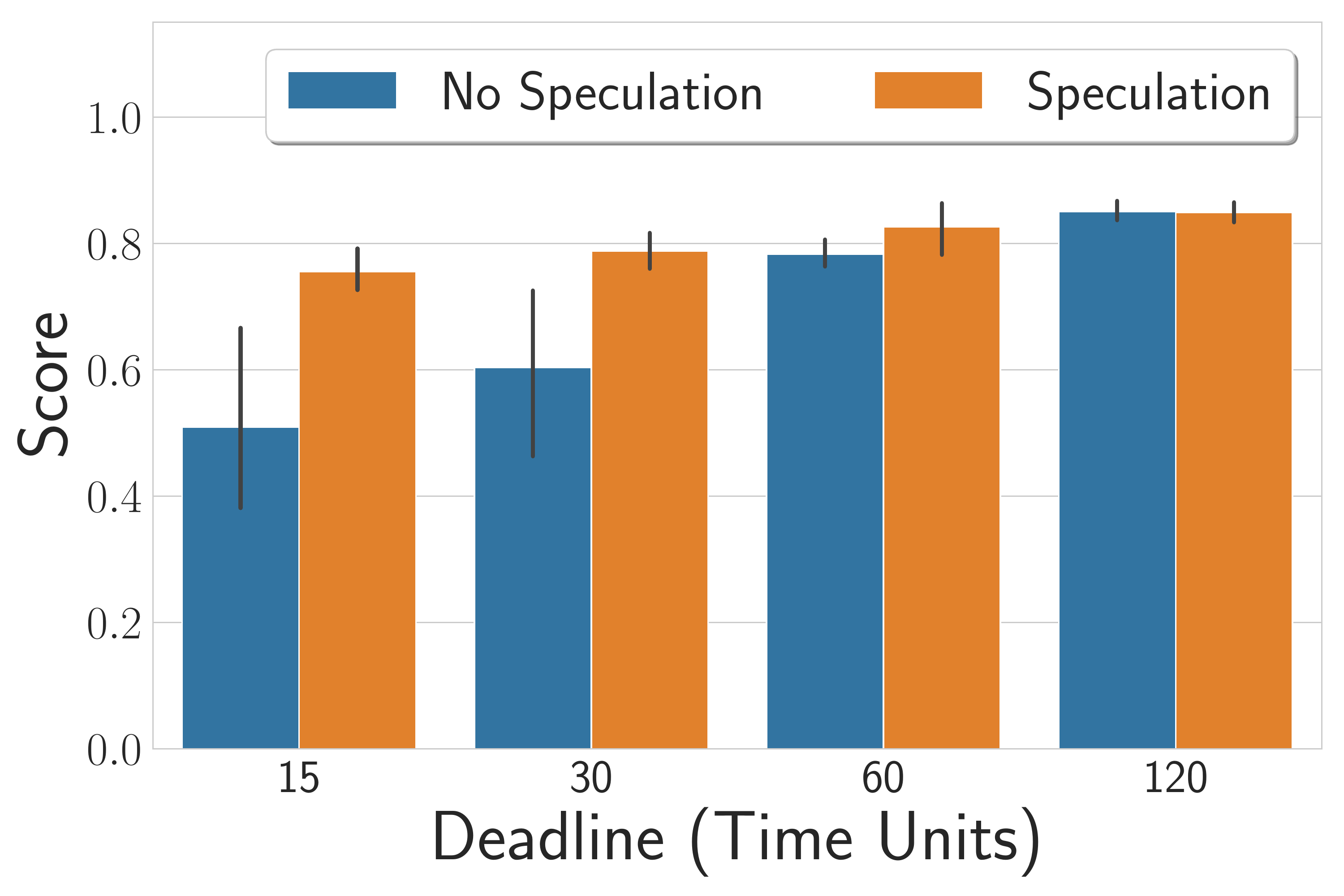}
        \caption{Score (16 atoms)}
    \end{subfigure}
    \hfill  
    \begin{subfigure}{0.24\textwidth}
        \includegraphics[width=1\linewidth]{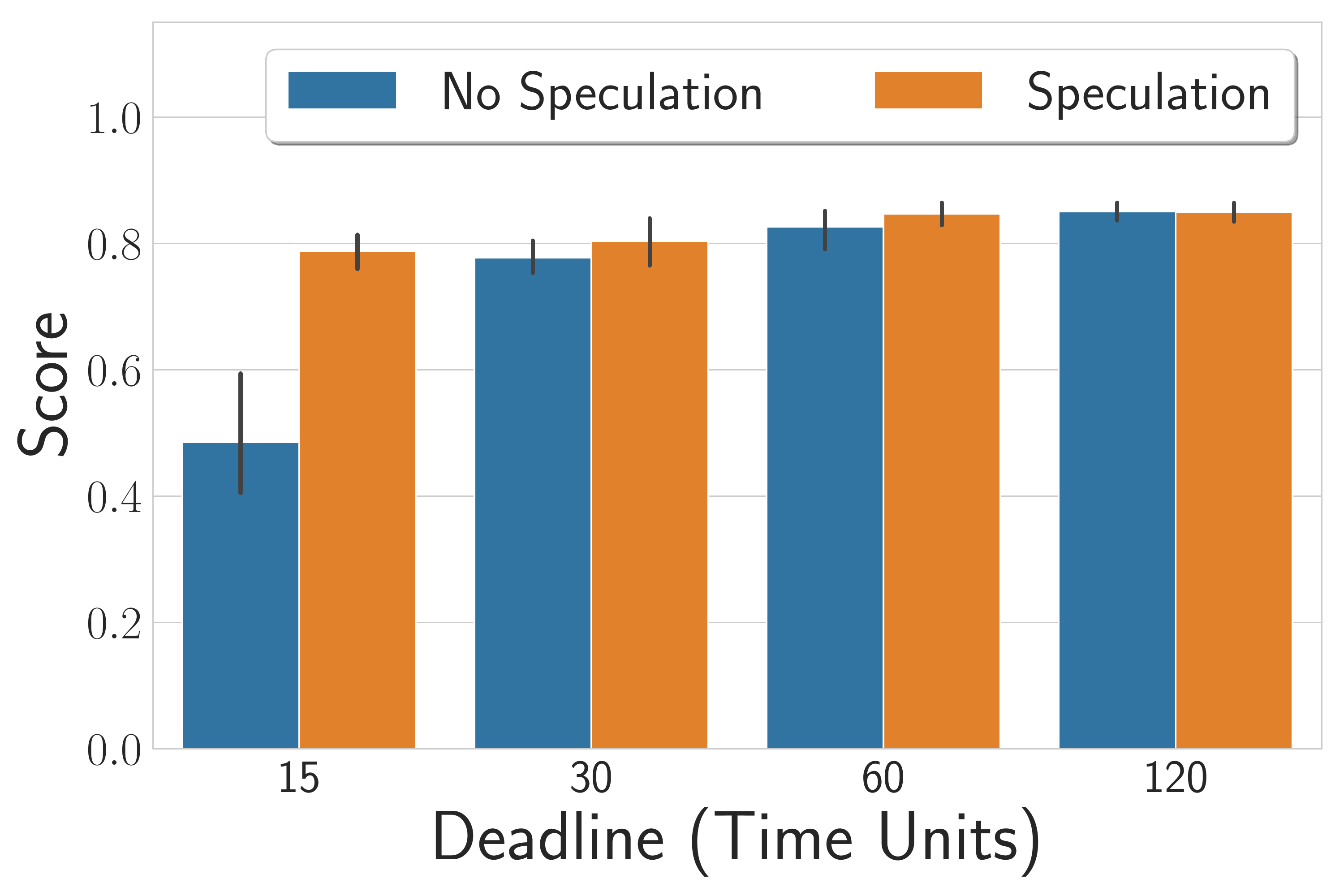}%
        \caption{Score (32 atoms)}
    \end{subfigure}
\caption{\textit{Speculative Evaluation}: Score and number of trials evaluated for various cluster sizes and deadlines. Adding a speculative evaluation mechanism can dramatically increase experiment performance, though the benefit decreases at larger scales.}
\label{fig:speculative}
\end{figure*}

In this section, we provide the following:
\begin{enumerate}
    \item an ablative study on synthetic workloads over the various design decisions of \hypersched{} 
    \item sensitivity experiments on synthetic workloads to evaluate \hypersched{}'s robustness
    \item demonstrate significant improvements over ASHA \cite{asha} with end-to-end time-constrained experiments for various deep learning training workloads. 
\end{enumerate}

\begin{figure}
    \centering
        \centering
        \includegraphics[width=.2\textwidth]{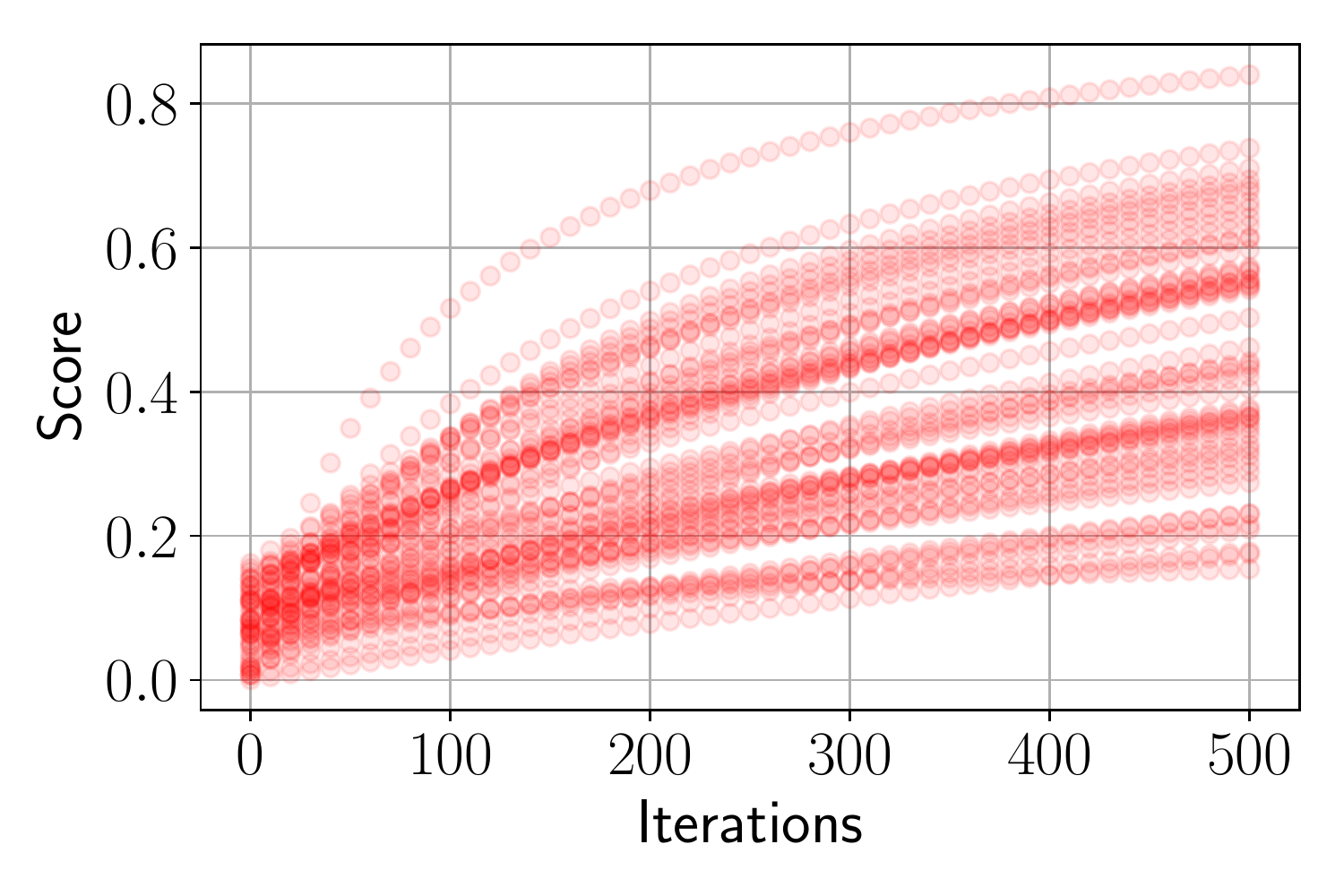}
  \caption{Score distribution of 50 sampled hyperparameter configurations of the simulated training function. This shows that there is fair spread of performance in a best effort to represent a realistic model tuning experiment.}
  \label{fig:optimus-score}
\end{figure} 
We run ablation and sensitivity experiments with evaluations of a simulated model training function ranging from 0 to 1: 
$$\text{Score(k)} = \Big(2 - (\frac{1}{0.01 b_0 k + 0.1b_1 + 0.5} + 0.01b_2)\Big)/ 2$$
where $b_1, b_2$ are drawn from a uniform distribution from 0 to 1, while $b_0$ is drawn from an exponential distribution with a scale factor of 0.1. 
This is a similar parametric model as used in \cite{optimus} and is used to represent \textbf{model accuracy}. We plot 50 samples of this parametric model and hyperparameter space in Figure~\ref{fig:optimus-score}.

All ablation and sensitivity experiments are done on an AWS c5.9xlarge which has 36 virtual cores. We simulate training with a deadline using \textit{time units} where each trial training step is simulated with $0.1$ time units unless otherwise stated. In simulation, each  \textit{time unit} corresponds to a second. Scaling is simulated by dividing the training step time by the scaling factor. All comparisons are evaluated over 5 seeds, which is necessary to produce the same ordering of hyperparameter configurations among comparisons during an experiment.

\subsubsection{Varying Scaling.}
\label{sec:scale-function}
In some experiments, we vary the scaling function of the underlying training function. This means that when more resources are allocated to the trial, the rate of training will increase by a factor dependent on the scaling function. We have the following scaling functions:

\begin{enumerate}
    \item \textit{LINEAR} scaling means that each training step time will be divided by the number of atoms allocated. 
    \item \textit{SQRT} scaling means that each training step time will be divided by the square root of atoms allocated. 
    \item \textit{NONE} scaling means that there is no change in behavior when more atoms are allocated to a trial.
\end{enumerate}

\begin{table}[h]
\begin{center}
\begin{tabular}{cccc}
\toprule
\textbf{Concurrent} & \multicolumn{3}{c}{\textbf{Step Time}}\\
\textbf{Trials}     & 0.01 & 0.1  & 0.5 \\
\midrule
8 & $1564 \pm 11$ &$ 287 \pm 1$ & 58 \\\hline
16 & $823 \pm 12$ & $287$ & 58  \\\hline
32 & $453 \pm 2$ & $284 \pm 1$ & 57  \\
\bottomrule
\end{tabular}
\end{center}
\caption{Average training steps of a trial taken by Tune, varying the amount of time one training step takes from 0 seconds to 0.5 seconds. These plots show that with a step time of 0.1 seconds, Tune is able to handle 32 concurrent trials with minimal scaling overhead. }
\label{table:tune-overhead}
\vspace{-8mm}
\end{table}

\textbf{Tune overheads}: We benchmark Tune to measure the amount of incurred overhead from the underlying execution framework. In Table~\ref{table:tune-overhead}, we provide the average number of steps per trial with different deadlines and different amounts of parallelism. We see that comparing 8 trials in parallel to 16 trials in parallel, the overhead becomes negligible when the trial steps are approximately 100ms. Startup overhead for Tune is around 1.5 seconds.

\subsection{Ablation Study}

In this section, we perform an ablative study of the various design aspects of \hypersched{}. Specifically, we aim to answer the following questions:
\begin{enumerate}
    \item What are the benefits of speculative evaluation?
    \item How robust is \hypersched{}'s adaptive entrance policy compared to a fixed entrance policy?
    \item How much benefit does dynamic resource allocation provide?
    \item How much benefit do the scaling and overhead information provide?
\end{enumerate}

\hypersched{} improves upon ASHA by limiting the number of trials that enter the system while reducing the amount of compute time spent on unpromising trials. This is done with two mechanisms: speculative evaluation and an entrance policy. Additionally, \hs{} dynamically allocates resources to the best-performing trials in a \textit{scale-aware} manner.

\begin{figure*}[ht]
    \centering
    \begin{subfigure}[t]{0.35\textwidth}
        \centering
        \includegraphics[width=\textwidth]{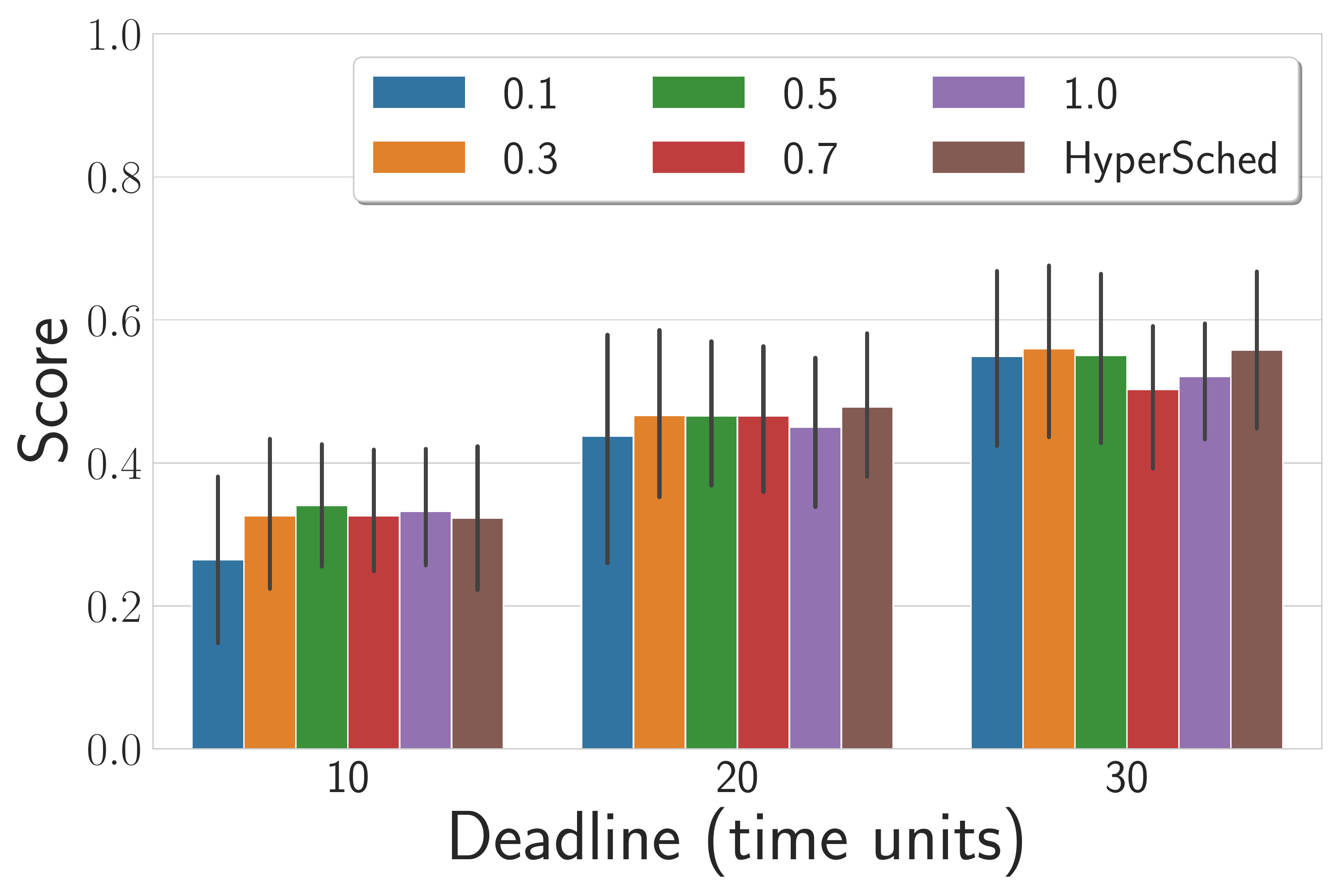}
        \caption{4 Atoms}\label{fig:exploration-4}
    \end{subfigure}
  \begin{subfigure}[t]{0.35\textwidth}
        \centering
        \includegraphics[width=\textwidth]{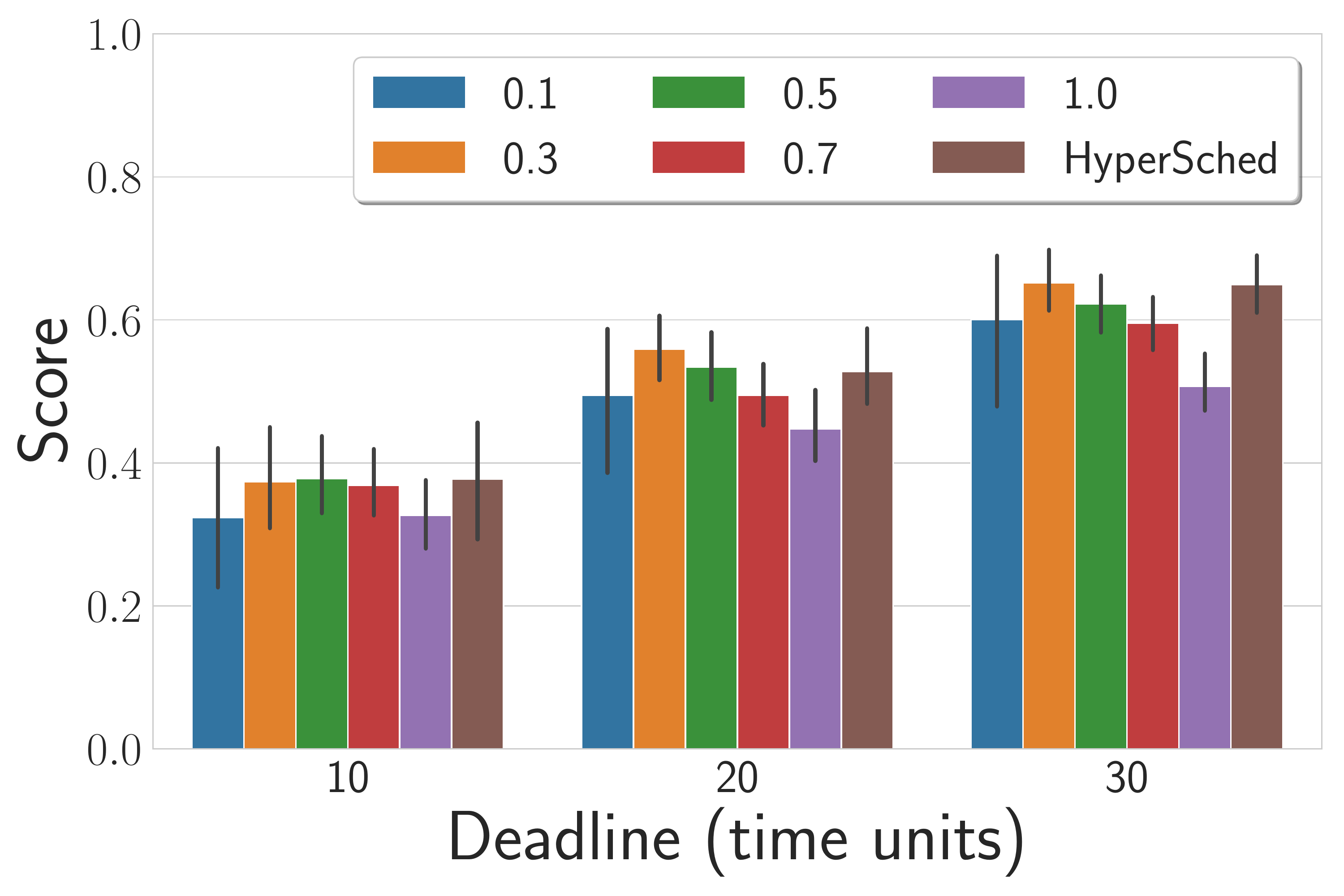}
        \caption{16 Atoms}\label{fig:exploration-16}
    \end{subfigure}
  \caption{\textit{Entrance Policy}: We compare the \hypersched{} entrance policy to different exploration policies that use a fixed "exploration time fraction", which represent the fraction of the time allocation when new trials are launched. After fraction$*T$ time has passed, no new trials will be started, and existing trials will be exploited. In this experiment, we evaluate the synthetic model training function on multiple deadlines and cluster sizes. We see that the entrance policy is quite robust, though it is not optimal.}
  \label{fig:exploration}
\end{figure*}

\textbf{Benefit of Speculative Evaluation:}  In Figure~\ref{fig:speculative}, we compared the performance of \HyperSched{} with and without speculative evaluation. Without speculative evaluation, trials are paused if they arrive early at a rung. The standard entrance policy is used. Our results show that speculative evaluation provides a significant performance increase. We use clusters of 4, 8, 16, and 32 resource atoms, and for each cluster, we evaluate the workload with deadlines of 15, 30, 60, and 120 time units. A one-atom trial step is simulated to take 0.1 time units and will decrease linearly with the number of atoms allocated. Each trial can take a maximum of 500 steps ($R$). As shown in Fig.~\ref{fig:speculative}, speculative evaluation can improve performance in simulation by up to 50\%. We also observe that the difference becomes less pronounced as the duration increases and as the cluster size increases. This is because at larger cluster sizes, more trials reach the first evaluation point (rung) faster, letting trials to progress without needing to wait. At longer deadlines, the fraction of time spent on waiting for trials to reach the first rung also diminishes, thereby diminishing the benefit of speculative evaluation.

\textbf{Benefit of entrance policy:} In Figure~\ref{fig:exploration}, we evaluate the effectiveness of the entrance policy mechanism. We compare our policy to a fixed policy that specifies a fraction of the time allocation for exploration. After this fraction has passed, no new trials will be launched. In smaller fractions (i.e., 0.1), we see degraded exploration and sub-par exploitation, while the larger fractions (i.e., 1.0) results in an inability to exploit the current running trials. Our experiments show that across 5 seeds, our heuristic-based entrance policy is robust to the length of the deadline (10, 20, 30 time units) and size of clusters (4 atoms, 16 atoms). \hypersched{} effectively exploits long-running jobs by limiting the number of jobs executed and often performs nearly as well as the best fixed-policy. 

\begin{figure}[t]
    \centering
    \begin{subfigure}{0.23\textwidth}
        \centering
        \includegraphics[width=\textwidth]{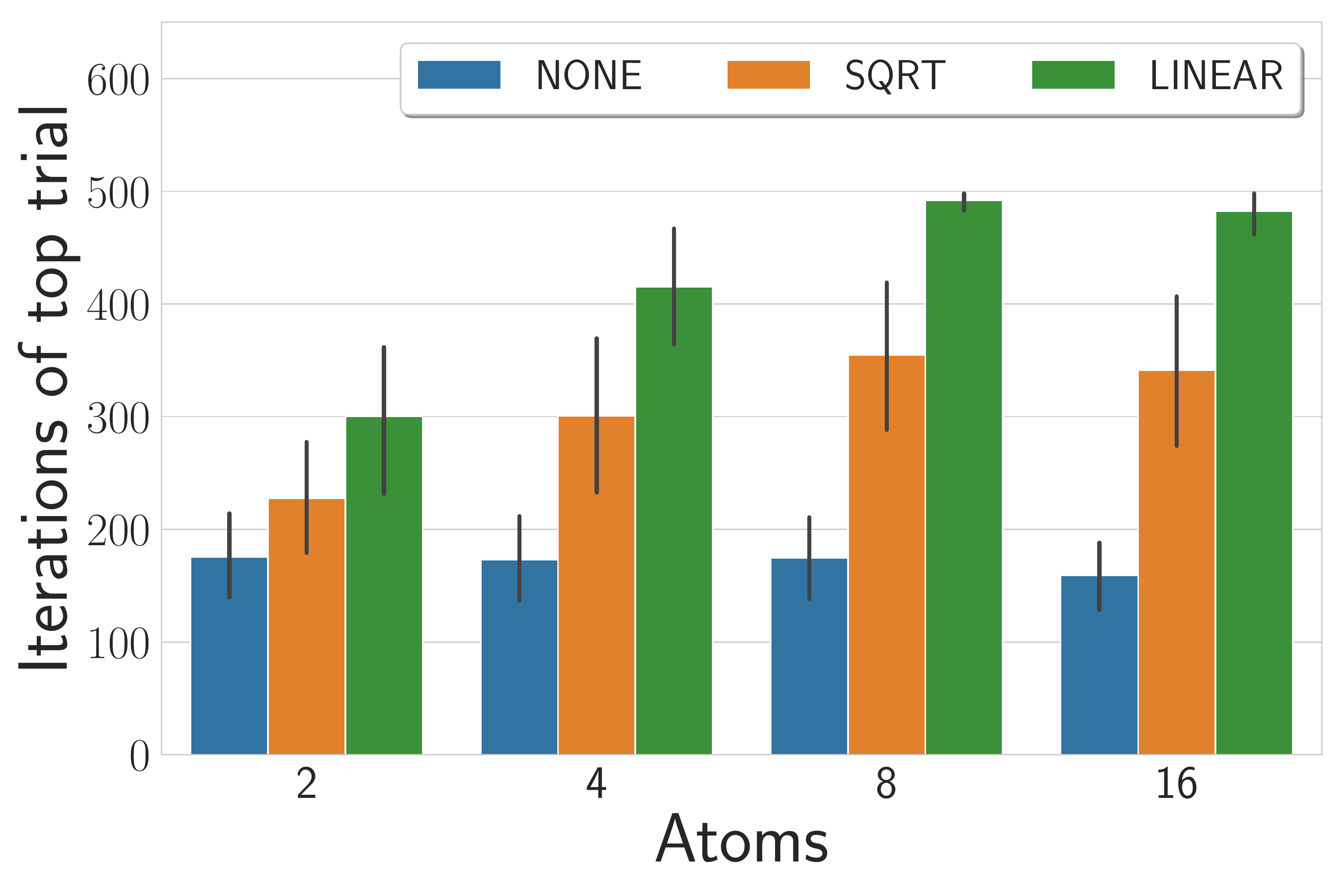}
        \caption{Iterations on Top Trial.}
    \label{fig:dra-iter}
    \end{subfigure}
    \begin{subfigure}{0.23\textwidth}
        \centering
        \includegraphics[width=\textwidth]{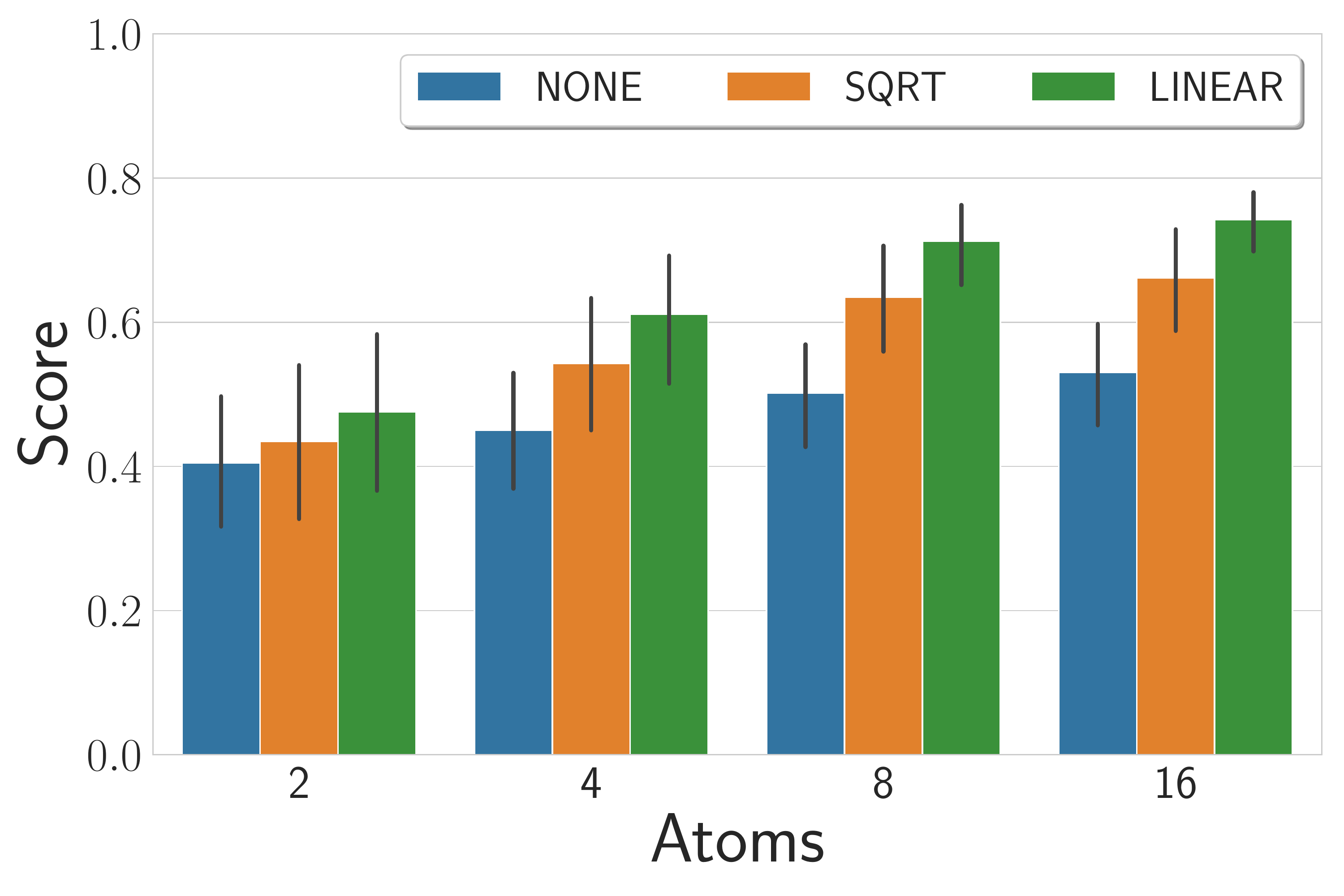}
        \caption{Score of top trial}\label{fig:dra-score}
    \end{subfigure}
  \caption{\small \textit{Dynamic Resource Allocation}: On a synthetic training function with various scaling properties, \Hypersched{} is able to effectively allocate resources and increase top job score across different cluster sizes and scaling properties. The deadline was 10 time units.}
  \label{fig:dra}
\end{figure}

\textbf{Benefit of Dynamic Resource Allocation}
In Figure~\ref{fig:dra}, we show the benefits of being able to scale jobs through various degrees of scalability - None, square-root scalability, and linear scalability (see Section~\ref{sec:scale-function} for an explanation of scaling functions). We show that across 4 different levels of parallelism (2, 4, 8, 16 atoms), \HyperSched{} increase performance by allocating more resources to the top job.

\begin{figure}
    \begin{subfigure}[t]{0.23\textwidth}
        \includegraphics[width=1\linewidth]{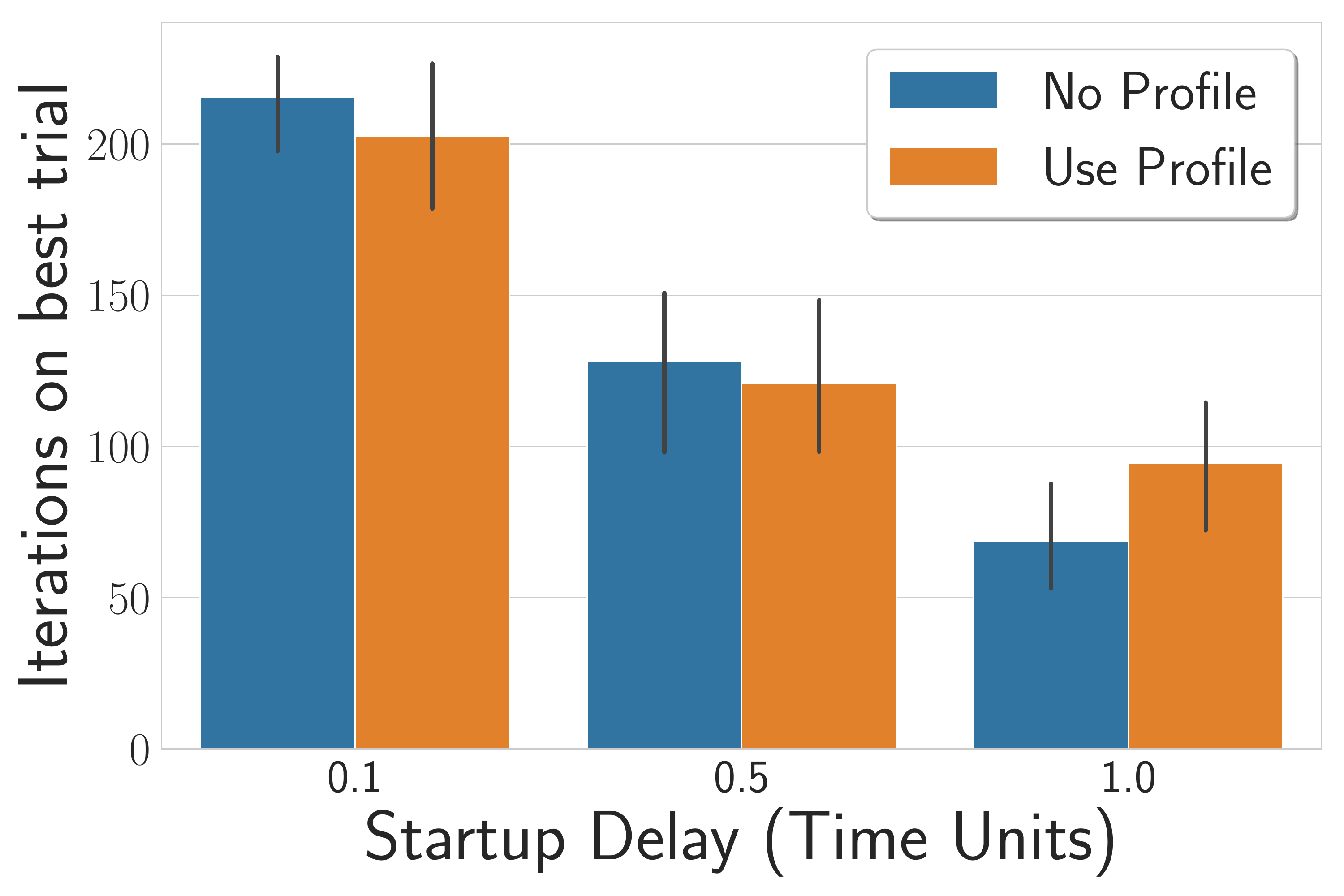}
        \caption{Iterations on Top Trial }\label{fig:a}
    \end{subfigure}
    \begin{subfigure}[t]{0.23\textwidth}
        \includegraphics[width=1\linewidth]{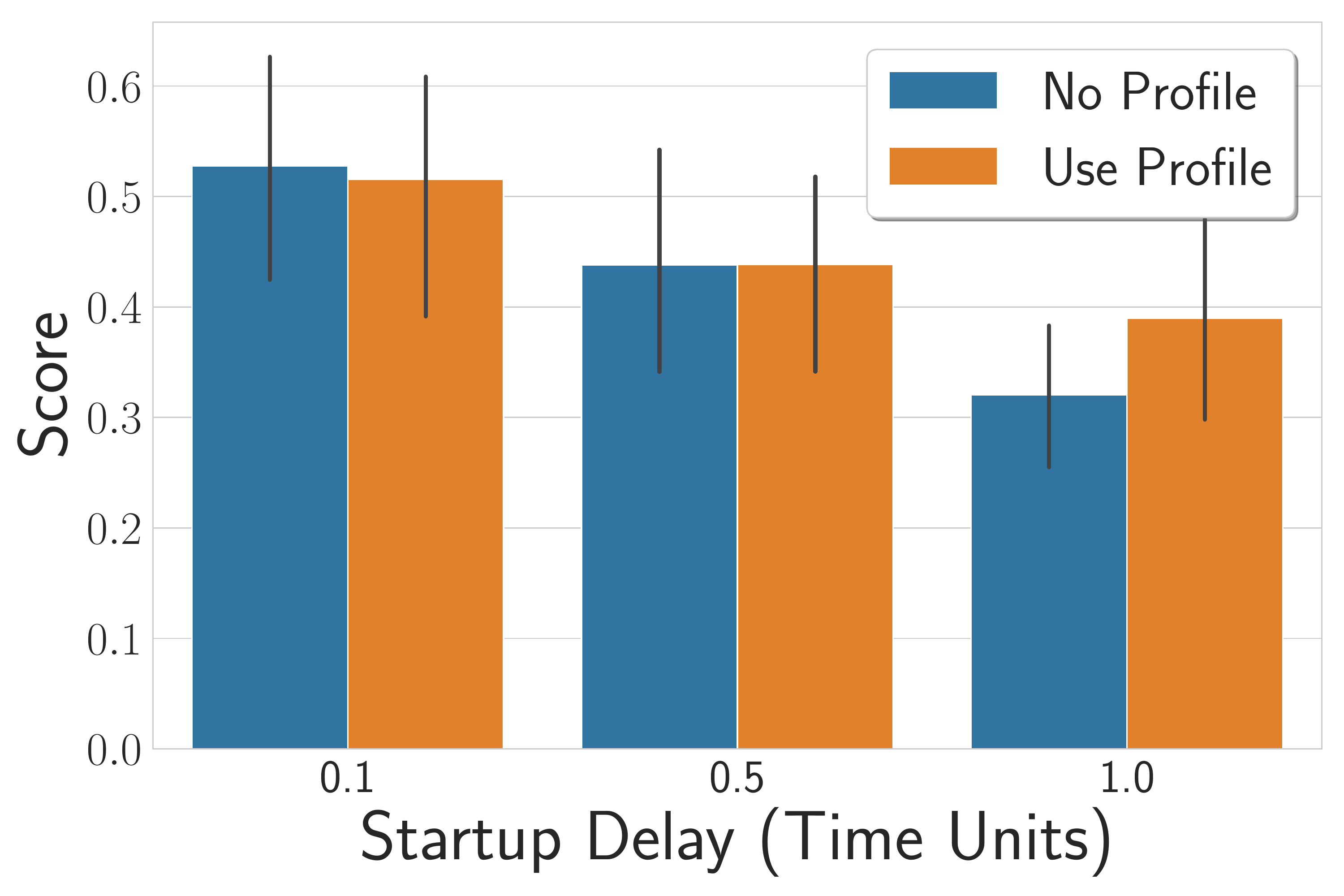}%
        \caption{Top Trial Score}\label{fig:b}
    \end{subfigure}
\caption{\small \textit{Profiling System Behavior}: As the startup delay for resizing and launching new trials increases, the benefit of system profiling becomes much more prominent. Total atoms was set to 16, and the deadline was 10 time units.}
\label{fig:overhead}
\end{figure}

\textbf{Benefit of Profiling System Behavior}
In Figure~\ref{fig:overhead}, we show that not providing system profile information to \hypersched{} can degrade the performance of the scheduling. In this experiment, we evaluate the performance of \hypersched{} when we blind-fold two profiling parameters: overhead tracking and scale-awareness. Specifically, we set the scaling function of the simulated trial to be a square root scaling function, and we toggle \hypersched{} to either internally use a linear scaling function without overhead tracking ("No Profile") or a square root scaling function with overhead tracking ("No Profile"). The use of the linear scaling function internally when the actual scaling is a square-root scaling can cause the scheduler to be too optimistic. We also vary the overhead for starting and resizing a trial ("startup delay") as a percentage of the overall deadline from $0.1 (1\%), 0.5 (5\%), 1.0 (10\%)$. We note that when the delay is smaller, profiling does not provide a significant difference. However, it can make a large difference in performance if the overhead deviates from the ideal (0 startup delay). This is due to the scheduler making sub-optimal decisions upon adaptive allocation. 

\subsection{Sensitivity Experiments}
\begin{figure}[h]
    \centering
    \begin{subfigure}[t]{0.23\textwidth}
        \centering
        \includegraphics[width=\textwidth]{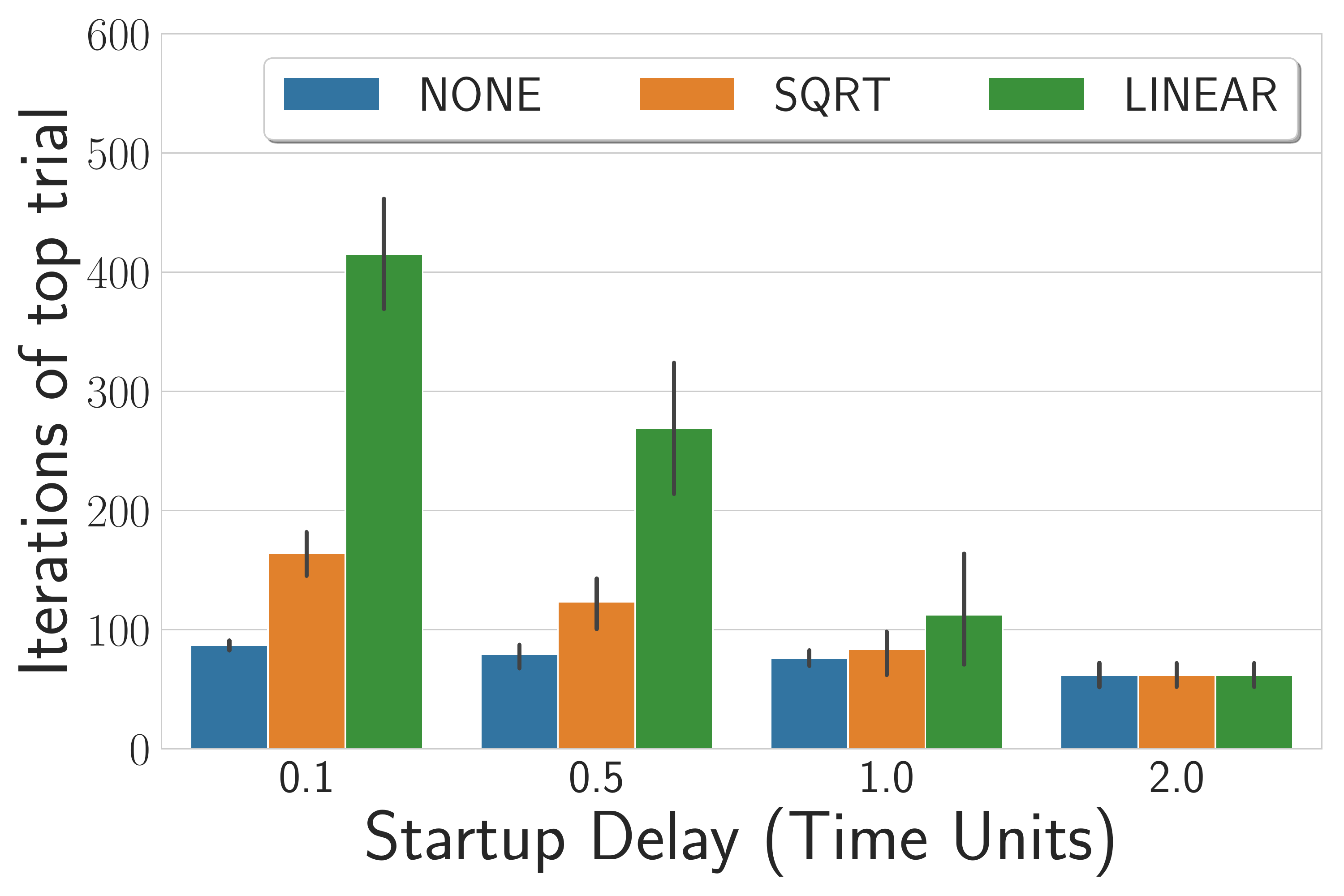}
    \caption{Scaling vs Iteration.}
    \label{fig:overhead-sensitivity-iter}
    \end{subfigure}
    \begin{subfigure}[t]{0.23\textwidth}
        \centering
        \includegraphics[width=\textwidth]{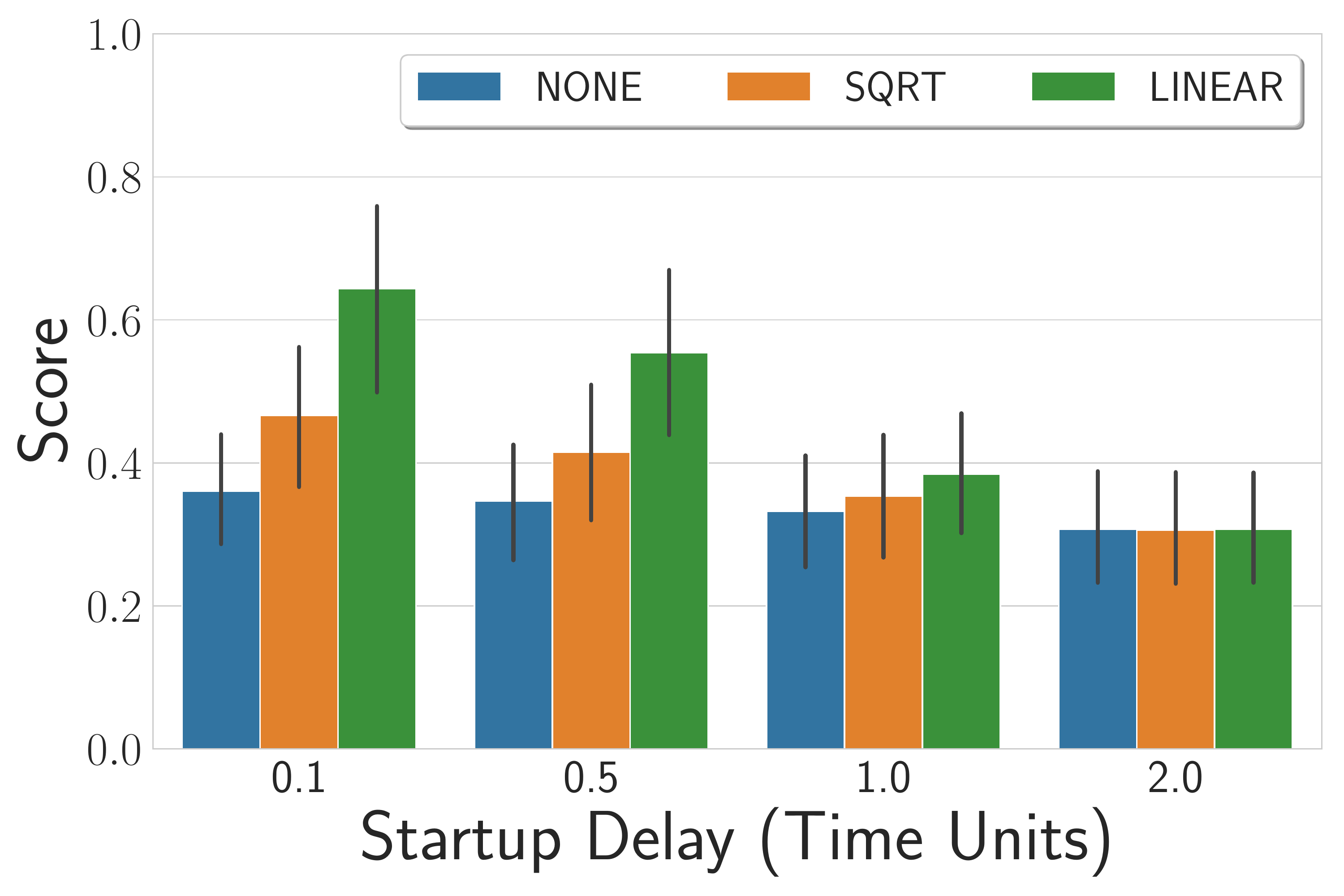}
        \caption{Scaling vs Score}\label{fig:overhead-sensitivity-acc}
    \end{subfigure}
   \caption{\textit{Sensitivity to scalability:} Performance of \hypersched{} on a simulated training function with linear, square root, and no scaling, varying the startup/resizing overhead. Increasing trial overhead can have significant effect on the amount of progress that \hypersched{} will make, depending on scaling behavior. The resource allocation is set to 20 time units and 4 atoms. }
  \label{fig:overhead-sensitivity}
\end{figure}

In Figure~\ref{fig:overhead-sensitivity}, we show the effects of increasing the overhead of starting or resizing on a simulated training by measuring the throughput (iterations / time-unit) and score of the top trial at deadline across various scaling behaviors. See Section~\ref{sec:scale-function} for an explanation of scaling functions "LINEAR", "SQRT", "NONE". We emulate the overhead of instantiating a large model in a realistic training workload by forcing the simulation to sleep right after resizing or starting. We range this overhead ("startup delay") as a percentage of the overall deadline from 0.1 $(1\%)$, 0.5 $(5\%)$, 1.0 $(10\%)$, 2.0 $(20\%)$ and show that the overhead will affect the performance of \hypersched{}, but \hypersched{} is still generally able to improve performance above the baseline of no scaling. At a large enough startup delay, \hypersched{} will not be able to provide any benefit but also does not perform worse than the baseline.

\subsection{Model-Training} 
\label{sec:data-parallel}
In this section, we compare \hypersched{} to ASHA on different space-time allocations and different deep learning models. 

\begin{table}[h]
\small
\begin{center}
\begin{tabular}{c|c}
\toprule
\textbf{Hyperparameter} & \textbf{Space} \\
\midrule
Learning Rate & 1e-4, 5e-4, 1e-3, 5e-3,  \\
  &   0.01, 0.05, 0.1, 0.5, 1 \\
\midrule
Weight Decay & 0.0001, 0.0005, 0.001, 0.005  \\
\midrule
Momentum & 0.9, 0.95, 0.99, 0.997 \\
\bottomrule
\end{tabular}
\end{center}
\caption{Data-Parallel training hyperparameter space for vision models. Each trial samples randomly from each value.}
\label{table:cifarspace}
\vspace{-6mm}
\end{table}

All data parallel training experiments (Sections~\ref{sec:hypersched-time} and \ref{sec:hypersched-models}) for CIFAR10 use the same hyperparameter space as shown in Table~\ref{table:cifarspace}. Samples from the hyperparameter space choose one value randomly from each list per hyperparameter. In all CIFAR10 experiments, a standard SGD optimizer is used. The learning rate decays by 0.2 at 60 epochs, 120 epochs, and 160 epochs. The batch size increases linearly with the number of GPUs, starting with a batch size of 128 for one atom. One atom is set to 1 GPU. We use model architectures and data transformations provided in a popular online implementation \cite{pytorchcifar}. We use PyTorch's framework specific utilities for checkpoint/restore along with Pytorch's data-parallel synchronization between updates. Note that we use a different setup for Section~\ref{sec:babi}.

In all experiments, ASHA and \hypersched{} share many of the same parameters. For both algorithms, we set the same values for $r$, the minimum resource allocation (where in our setting resource is epochs) and $\eta$, the reduction factor. We set the minimum resource allocation to be 2 epochs of training which results in roughly 60 seconds of execution. The max resource allocation (epochs) is dependent on the experiment.
Our reduction factor for both algorithms is set at 4, which is a standard value \cite{asha}.

\textbf{Large Batch Training:} Increasing the batch size by itself does not increase the rate of learning. for the rate of model updates does not change. Therefore, we follow the techniques commonly used in the literature to achieve scaling benefits with larger batch sizes \cite{gholami, goyal2017accurate}. When resizing, we scale the learning rate by a factor that is square-root proportional to the batch size. This is enabled by the larger batch size and allows each gradient update to provide more training progress. Further, because increasing the learning rate dramatically at once will destabilize learning, we use a \textit{ramp-up} period for the learning rate \cite{he2016deep}, gradually increasing it by no more than 10\% of the previous value per epoch. This ramp-up occurs every time the models resize.

\subsubsection{\hypersched{} performance across time}
\label{sec:hypersched-time}
\begin{figure}[t]
    \centering
    \begin{subfigure}{0.23\textwidth}
        \centering
        \includegraphics[width=\textwidth]{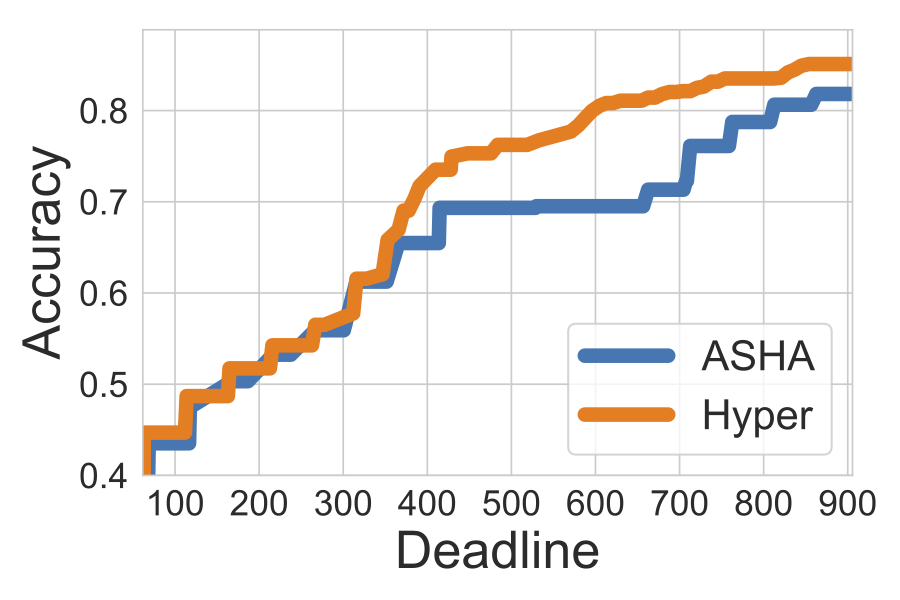}
    \caption{Deadline = 900 s}
    \label{fig:cifar-900}
    \end{subfigure}
    \begin{subfigure}{0.23\textwidth}
        \centering
        \includegraphics[width=\textwidth]{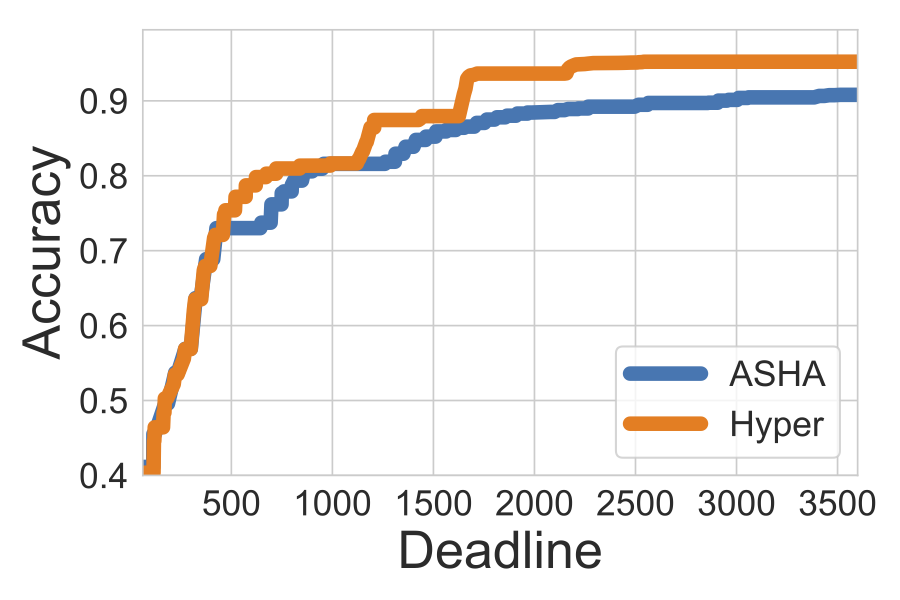}
        \caption{Deadline = 3600 s}\label{fig:cifar-1800}
    \end{subfigure}
    \caption{Max accuracy over time for \hypersched{} and ASHA with training deadlines of 900 and 3600 seconds. \hypersched{} is able to consistently select and train configurations to higher accuracies than ASHA. The experiment is given 8 atoms.}
\label{fig:pytorch30}
\end{figure}

\begin{table}
\begin{tabular}{|c|c|c|c|c|}
\hline
\textbf{Model}                 & \textbf{1 GPU} & \textbf{2 GPUs} & \textbf{4 GPUs} & \textbf{8 GPUs} \\ \hline
\textbf{ResNet18}&1.00x&1.85x&3.43x&5.40x \\ \hline
\textbf{ResNet50}&1.00x&1.87x&3.42x&5.60x \\ \hline
\textbf{ResNet101}&1.00x&1.89x&3.63x&6.67x \\ \hline
\textbf{ShuffleNet V2}&1.00x&1.92x&3.61x&6.08x \\ \hline
\end{tabular}
\caption{Resource scaling profiles of different models, measured as normalized throughput of models with different resource allocations. All models exhibit a sub-linear scaling with increased resource allocation. We evaluate this on a single node with 8 NVIDIA V100s.}
\label{table:scaling}
\vspace{-8mm}
\end{table}
In this experiment, we train ResNet50 models on CIFAR10 on a p3.16xlarge instance which has 8 NVIDIA V100 GPUs.
$r$ is 2 epochs, and $R$ is 200 epochs. We use 8 atoms total, each atom representing one GPU. We evaluate this setup on three seeds. The goal of this experiment is to compare the performance of \hypersched{} on different deadlines: 900 seconds, 1800 seconds, and 3600 seconds.


As seen in Table~\ref{table:resnet-time}, \hypersched{} is able to outperform ASHA on a variety of different deadlines. As the deadline increases, ASHA performs better and better since models gradually converge and have diminishing increases in accuracy.

\begin{table}[]
\begin{tabular}{|c|c|c|c|c|}
\hline
\textbf{Deadline (s)}      & \textbf{Scheduler} & \textbf{Max Accuracy} & \textbf{Num. of trials} \\ \hline
\multirow{2}{*}{900}       & ASHA           & 0.833 $\pm$ 0.017 & 35 $\pm$ 3     \\ \cline{2-4} 
                           & \HyperSched{}     & \textbf{0.879 $\pm$ 0.017} & 14       \\ \hline
\multirow{2}{*}{1800}      & ASHA           & 0.886 $\pm$  0.016  & 51 $\pm$ 5       \\ \cline{2-4} 
                           & \HyperSched{}     & \textbf{0.916 $\pm$ 0.029} & 20 $\pm$ 1       \\ \hline
\multirow{2}{*}{3600 (1h)} & ASHA           & 0.908 $\pm$ 0.003 & 90 $\pm$ 15       \\ \cline{2-4} 
                           & \HyperSched{}     & \textbf{0.938 $\pm$ 0.009} & 33 $\pm$ 4       \\ \hline
\end{tabular}
\caption{Maximum accuracy at deadline comparing ASHA and \HyperSched{} on training ResNet-50 on CIFAR. We see that across the board, ASHA is able to evaluate more trials than \hypersched{}. However, \hypersched{} is always able to achieve a higher top accuracy at deadline. } 
\label{table:resnet-time}
\vspace{-8mm}
\end{table}

For reference, a ResNet50 model with \textbf{pre-tuned} hyperparameters on CIFAR10 on a single V100 (assuming 100 epochs and 40 seconds per epoch) will take 1 hour to train to 93\% accuracy \cite{pytorchcifar}. \hypersched{} can reach over $87\%$ accuracy, evaluating 14 models in $1/4$ of the time (2x the equivalent amount of resource-time).

\subsubsection{\hypersched{} performance across models}
\label{sec:hypersched-models}
We show the robustness of \hypersched{} across various models in this section. We train ShuffleNet v2, ResNet18, ResNet50 and ResNet101 individually on CIFAR10 on AWS p3.16xlarge instances (each with 8 V100 GPUs) with a 900 second deadline ($T$). 
Each trial is trained for at most 100 epochs, corresponding to $R$. Each trial is also trained for a minimum of 2 epochs, corresponding to $r$.
These models are representative of a wide range of compute requirements with ResNet101 requiring the most compute cycles per image.  For each model, we also evaluate performance on different amounts of parallelism - 4 GPUs and 8 GPUs. For \hypersched{}, atoms are set at 1 GPU per atom. Both ASHA and \hypersched{} use a reduction factor of $4$.
Each configuration is evaluated on 3 seeds to seed the random hyperparameter samples drawn from the same space as shown in Table~\ref{table:cifarspace}, and both ASHA and \hypersched{} are evaluated on the same sequence of trials for each seed. 

We first benchmark how much benefit changing the resource allocation for a particular model provides. Table ~\ref{table:scaling} presents the sub-linear scaling exhibited by four popular deep learning models when the batch-sizes are scaled in proportion to the number of GPUs allocated. All models follow a sub-linear scaling pattern, where allocating an increasing amount of GPUs has diminishing returns due to overheads of distributing computation across GPUs. This motivates the \textit{scale-aware} design of \hypersched{}, which allows it to evaluate the potential benefit from resizing trials.

We see in Table~\ref{table:modelatoms} that across various neural network architectures, \hypersched{} is able to consistently outperform ASHA across different resource configurations. For more resource intensive models like ResNet101, \HyperSched{} exhibits larger gains over ASHA - up to 10\% more accuracy points - because of its ability to dynamically reallocate resources in a deadline-aware fashion. For smaller models like ResNet18, the 900 second deadline is long enough for the model to converge even without resource re-allocation, and thus the gap between ASHA and \hypersched{} is smaller. However, at shorter deadlines, ResNet18 would demonstrate greater gains from using \hypersched{} than ASHA.

\begin{table}[]
\begin{tabular}{|c|c|c|c|}
\hline
\textbf{Model}                 & \textbf{GPUs} & \textbf{ASHA} & \textbf{\hypersched{}} \\ \hline
\multirow{2}{*}{ResNet 101} & 4              & 0.70 $\pm$ 0.05 & \textbf{0.73 $\pm$ 0.11}       \\ \cline{2-4} 
                               & 8              & 0.71 $\pm$ 0.05 & \textbf{0.81 $\pm$ 0.03}       \\ \hline
\multirow{2}{*}{ResNet 50}    & 4              & 0.79 $\pm$ 0.05 & \textbf{0.87 $\pm$ 0.03}       \\ \cline{2-4} 
                               & 8              & 0.83 $\pm$ 0.01 & \textbf{0.89 $\pm$ 0.02}       \\ \hline
\multirow{2}{*}{ResNet 18}    & 4              & 0.90 $\pm$ 0.00 & \textbf{0.93 $\pm$ 0.00}       \\ \cline{2-4} 
                               & 8              & 0.90 $\pm$ 0.00 & \textbf{0.93 $\pm$ 0.02}       \\ \hline
\multirow{2}{*}{ShuffleNet V2}     & 4              & 0.80 $\pm$ 0.02 & \textbf{0.89 $\pm$ 0.02}       \\ \cline{2-4} 
                               & 8              & 0.82 $\pm$ 0.02 & \textbf{0.90 $\pm$ 0.01}       \\ \hline
\end{tabular}
\caption{Accuracy of multiple models on CIFAR10 across 3 runs. The deadline here is set to 15 minutes (900 seconds). For each model, two different resource configurations are evaluated (4 and 8 GPUs) to demonstrate \hypersched{}'s ability to adapt to varying resource configurations.}
\label{table:modelatoms}
\vspace{-8mm}
\end{table}

\subsubsection{Beyond vision models}
\label{sec:babi}
\begin{figure}[ht]
    \centering
    \begin{subfigure}{.23\textwidth}
        \centering
        \includegraphics[width=\textwidth]{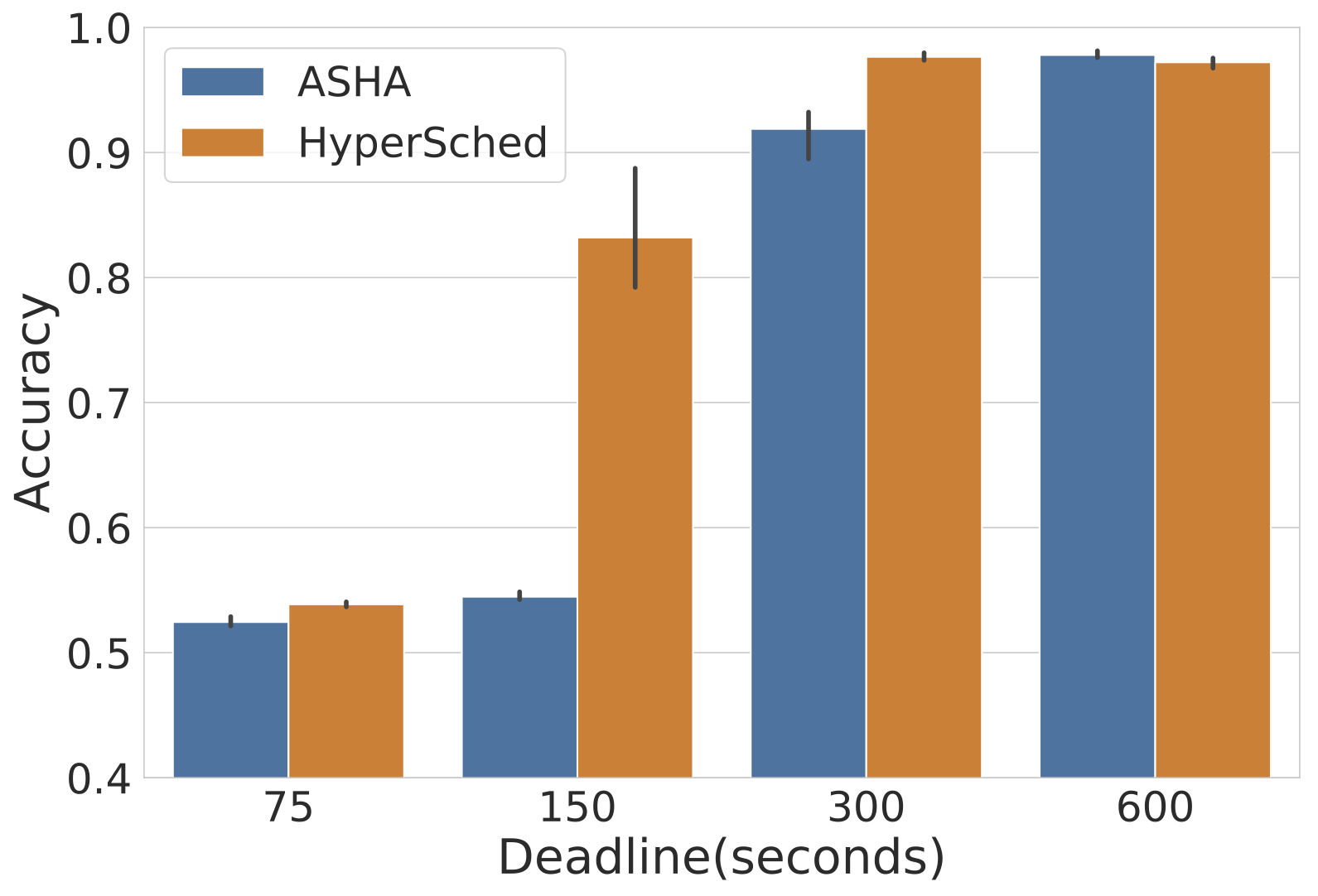}
        \caption{MemN2N on bAbI}
        \label{fig:deadlinefinalacc:babi}
    \end{subfigure}
    ~
    \begin{subfigure}{.23\textwidth}
        \centering
        \includegraphics[width=\textwidth]{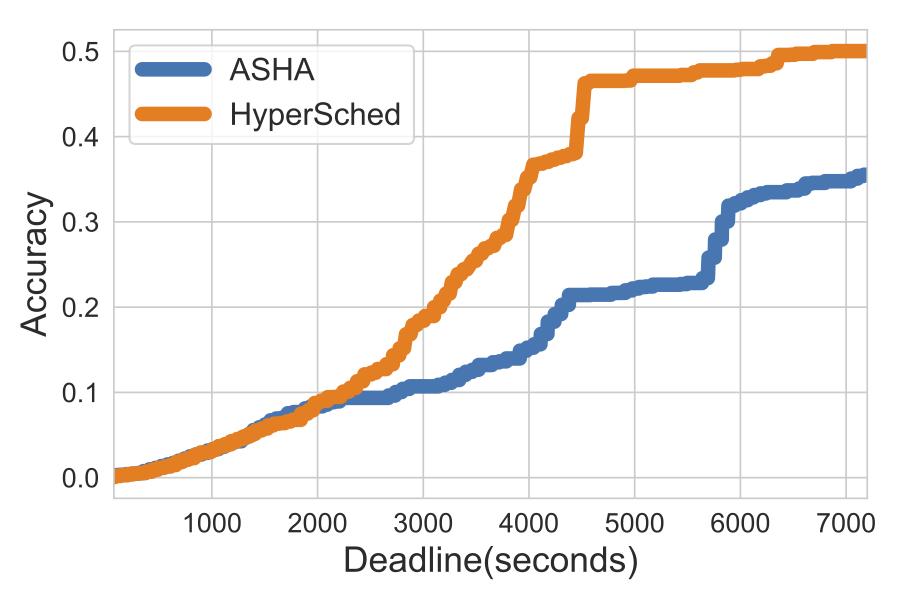}
        \caption{ResNet50 on Imagenet}
        \label{fig:deadlinefinalacc:imagenet}
    \end{subfigure}
    \caption{Figure~\ref{fig:deadlinefinalacc:babi} shows the final accuracy of the best performing job using \hypersched{} vs ASHA to train MemNN, a language model, on various deadlines. \hypersched{} obtains a higher final accuracy in shorter deadlines. Figure~\ref{fig:deadlinefinalacc:imagenet} shows the max accuracy over time for training ResNet50 on ImageNet using ASHA and \hs{} on a 32-GPU cluster.}
\label{fig:deadlinefinalacc}
\vspace{-5mm}
\end{figure}

We also demonstrate the effectiveness of \hypersched{} beyond vision models and GPUs. We train a MemN2N \cite{memn2n} model on the bAbI \cite{babi} single supporting fact task for text reasoning and understanding. This model is trained purely on CPUs and the experiment is run on a c5.9xlarge AWS instance with 36 virtual cores. Each atom corresponds to one CPU core. Fig.~\ref{fig:deadlinefinalacc:babi} compares the performance of \hypersched{} against ASHA. The hyperparameter space consists of a uniform distribution $U(4, 128)$ for the LSTM state size, $U(32, 128)$ for embedding size, $U(0.1, 0.5)$ for the dropout rate in internal layers, and a choice of SGD optimizer among \{"rmsprop", "adam", "sgd"\}. \hypersched{} is able to identify and prioritize promising trials to achieve a higher maximum accuracy across various deadlines. Note that the benefit over ASHA diminishes in longer time horizons. For reference, a fine-tuned implementation reaches 98.6\% accuracy after 120 epochs. $r$ is set to $20$ epochs, and $R$ is $300$ epochs. 

\begin{table}[h]
\begin{tabular}{|c|c|c|c|c|}
\hline
\textbf{Deadline (s)}      & \textbf{Scheduler} & \textbf{Max Accuracy} & \textbf{Num. of trials} \\ \hline
\multirow{2}{*}{7200}       & ASHA             & 0.367 $\pm$ 0.014          & 86 $\pm$ 16     \\ \cline{2-4} 
                           & \HyperSched{}     & \textbf{0.542 $\pm$ 0.036} & 36 $\pm$ 1       \\ \hline
\end{tabular}
\caption{Max accuracy of the best performing job using \hypersched{} and ASHA training ResNet50 on ImageNet with deadline of 7200 on a 32-GPU cluster, averaged over 3 runs.}
\vspace{-5mm}
\label{table:pytorch-distributed}
\end{table}

\subsection{Multi-Node Distributed SGD}

In this section, we run a distributed experiment over 4 AWS p3.16xlarge instances, each with 8 NVIDIA V100 GPUs. The purpose of this experiment is to show that \hypersched{} maintains performance benefits even at larger scales.

We train ResNet50 on the ImageNet dataset with the same hyperparameter space and training configurations (i.e. learning rate decay, batch size) as in Section~\ref{sec:data-parallel}. We use NCCL as the distributed communication backend. We use a deadline of 2 hours (7200 seconds), and we set atoms to 2 GPUs (for a total of 16 atoms). To increase the rate of feedback, we set $r$ (minimum resource allocation parameter) to be 10 sub-epochs, where each "sub-epoch" is 100 gradient updates and 20 validation batches. We reduce the amount of resizing by enforcing atom allocations greater than 1 to only be multiples of 2 -- empirically, this allows us to avoid NCCL failures.

We see in Table~\ref{table:pytorch-distributed} and Figure~\ref{fig:deadlinefinalacc:imagenet} that even in a larger scale and in the distributed setting, \hypersched{} is able to outperform ASHA significantly. Initially, the best jobs in both ASHA and \hypersched{} progress at a similar rate. As the 7200 second deadline approaches, Hypersched prioritizes training the best performing trial. 

Note that this benchmark number is not highly optimized, unlike other existing ImageNet training results~\cite{goyal2017accurate} which use techniques such as mixed-precision training. Also, because \hypersched{} does not manage worker placement, the results of this experiment can be improved with better placement strategies (i.e., colocating training workers) along with an allocation strategy that is aware of the underlying topology.



\section{Discussion}

\subsubsection{Limitations and future work}
Below, we discuss the current limitations of \hypersched{} and avenues of future work. The current entrance policy can make decisions based on the epoch/iteration time given one atom, but this quantity may be overly pessimistic. One adjustment would be to assume that more resources will be allocated to a new promising job.
Another opportunity for improvement is to reduce inefficiencies in distributed settings. Colocating distributed workers for data parallel training can provide significant performance improvements at small GPU allocations (i.e., a model training on 2 GPUs on the same node vs on 2 separate nodes). A possible avenue for \hypersched{} would include the ability to optimize placement of atom allocations and factor in cluster fragmentation in the exploration policy. 

\subsection{Related work}
\subsubsection{Scheduling for Deep Learning Workloads}
There has been a recent proliferation of systems and frameworks that address
resource scheduling for DDL. \HyperSched{} is an \textit{application-level scheduler} and given the ability to drop jobs depending on performance, hence forgoing the consideration of job completion time, fairness, or starvation.

Tiresias \cite{tiresias} is a cluster scheduling framework that schedules and places
distributed deep learning jobs to reduce Job Completion Times (JCT) for distributed deep learning workloads without prior modeling of training nor scaling performance. 
Gandiva \cite{Gandiva} is a scheduling framework for deep learning which aims to both provide
early feedback to jobs and maximize cluster efficiency. Gandiva introduces
scheduling primitives (grow-shrink, migration, packing, and checkpoint-restore)
that are broadly applicable to any deep learning cluster scheduler. It is important to note that Tiresias and Gandiva can serve as underlying low-level schedulers for \HyperSched{} and that mechanisms such as packing and migration could improve the performance of \HyperSched{}. 

Optimus \cite{optimus} is a cluster scheduling framework designed to minimize job completion
times and dynamically schedules resources to jobs based on progress to minimize
average job completion times. The authors use domain knowledge to craft parametric performance models and training performance models which are fit on the fly. \Hypersched{} optimizes a different objective and aims to address the general case where domain knowledge of resource performance is not necessarily known, and training performance models are not given.

SLAQ \cite{slaq} is a cluster scheduling system that hosts multi-tenant machine learning
training jobs running on shared resources. It is quality driven and provides
scheduling policies for maximizing total loss reduction per resource
allocation and maximizing the minimum quality across a set of trials.
In the single-tenant case where all trials optimize a shared objective, we note that the SLAQ objectives can be detrimental to performance (where the worst trial with degenerate hyperparameters is given all of the resources). \Hypersched{} specifically does not address these objectives and aims to optimize a global objective across multiple trials under resource constraints.

\subsubsection{Hyperparameter Optimization.} There are numerous services and
algorithms for addressing hyperparameter optimization. Google Vizier~\cite{golovin2017google}
is a black-box service for hyperparameter search that can
prematurely terminate underperforming trials automatically and allows new
trials to start on the newly available resources. However, this framework does
not support dynamic reallocation of resources to an existing trial.

HyperDrive \cite{hyperdrive} is a hyperparameter exploration framework that
identifies "promising", "opportunistic", and "poor" configurations
of hyperparameters using a combination of probabilistic dynamic
model-based classification, early termination, and prioritized resource scheduling
to jointly optimize quality and cost. The HyperDrive
authors, however, discuss dynamic resource allocation in the context of pooling
resources for a collection of jobs, each with their own static resource
allocation. \HyperSched{} considers the case where individual jobs are growth-capable and is specifically concerned with time constraints.

Hyperband \cite{HyperBand} and ASHA \cite{asha} are hyperparameter evaluation algorithms that
allocate resources dynamically to training jobs. However, Hyperband's budget is one-dimensional, which does not translate effectively into the practical space-time dimensions of scheduling as not all jobs are linearly scalable.

\subsubsection{Cluster scheduling} 
Existing resource managers and schedulers such as Kubernetes are orthogonal to \HyperSched{}, as they do not operate at the application-level and can be run at a level underneath. \HyperSched{} jobs can be considered as one single allocation request in a larger scheduling framework. Other work such as TetriSched \cite{tetrisched} exposes a language for utility-based scheduling. This can be
compatible with \HyperSched{} providing trial state as a utility function for the scheduler to use. 

\section{Conclusion}
\label{sec:conclusion}
We present \hypersched{}, a novel application-level scheduler designed to maximize accuracy under time and resource constraints. By leveraging scale-aware dynamic resource allocation, \hypersched{} is able to quickly identify and accelerate trials with the most promising set of hyperparameters. Combined with deadline aware exploration-exploitation phases, \hypersched{} outperforms current state-of-the-art hyperparameter search algorithms. 

\section {Acknowledgements}
We thank our shepherd Srinivasan Parthasarathy and the anonymous reviewers for their valuable feedback and suggestions to improve this work. In addition to NSF CISE Expeditions Award CCF-1730628, this research is supported by gifts from Alibaba, Amazon Web Services, Ant Financial, CapitalOne, Ericsson, Facebook, Futurewei, Google, Intel, Microsoft, Nvidia, Scotiabank, Splunk and VMware.
%
\bibliographystyle{ACM-Reference-Format}
\bibliography{main}

%
\appendix

\end{document}